# Diffuse interstellar bands (DIB): co-planar doubly excited He and metal atoms embedded in Rydberg Matter

Leif Holmlid

Atmospheric Science, Department of Chemistry, University of Gothenburg, SE-412 96 Göteborg, Sweden. E-mail: holmlid@chem.gu.se

## Abstract

The interpretation of the more than 300 diffuse interstellar bands (DIBs) is one of the most long-standing problems in interstellar spectra since the two first bands were reported in 1921. We now predict the frequencies of 260 diffuse interstellar bands (DIBs) using the Rydberg Matter model we have developed previously. These transitions involve mainly He atoms, but other two-electron atoms like Ca and other metals can take part in the absorption processes. Approximately 70% of the total intensity of the DIBs is due to absorption in doubly excited states and 30% in singly excited He atoms. The doubly excited states are in inverted states while the He atoms are thermal. The possibilities to observe DIBs in the UV and NIR ranges are discussed and band positions are predicted.

## 1. Introduction

The first observation of what is now called diffuse interstellar bands (DIBs) was made by Heger (1921) as two “stationary” features from binary stars, thus due to absorptions in the ISM. Several pioneering studies of DIBs, giving tables of intensities and widths and other characteristics of the bands, were made by Herbig (1975, 1993, 1995, e.g.). Many suggestions have been made during the 90 years of study to explain the numerous DIBs: more than 300 DIBs are detected so far. Extensive reviews are published (Herbig 1995; Krelowski 1999) and several surveys observing new bands have been published (Jenniskens & Désert 1994; Tuairisg et al. 2000; Weselak et al. 2000; Galazutdinov et al. 2000). There has existed three main ideas to uncover the origin of the DIBs: 1) they can be divided into families of lines which correlate in intensity and thus may belong to the same carrier (e.g. the same type of molecule) (Krelowski & Walker 1987), 2) some groups of lines can be singled out since their intensities correlate with different atomic or molecular lines (Weselak et al. 2004), and 3) high resolution spectroscopy gives the band shapes and thus information on the formation processes (Galazutdinov et al. 2003). None of these principles has been highly successful. They are discussed relative to the DIBs formation by doubly excited states in RM in Holmlid (2008a). The suggestion of at least one of the families of DIBs is shown below to agree fairly well with the present theoretical description. The long-standing idea that the DIBs are due to large organic molecules like polyaromatic hydrocarbons (PAHs) has not been successful either. The most apparent problem with such a concept is that for large molecules, the absorption spectra will be very sensitive to the temperature and other environmental factors, which is well-known. This has been shown for example in molecular beams, free jets and low temperature gas matrices, as described in standard textbooks on spectroscopy (Hollas 1998, 2004). Since the intense DIBs are almost unchanged in their width and shape in many different astronomical objects even if their intensities change and thus probably also the temperature of the DIB carriers, they are not due to any complex molecules. Instead, DIBs are clearly due to atomic transitions that are not strongly influenced by the temperature or other environmental parameters like the UV flux density. With atomic transitions inside a phase which defines the environment as in the Rydberg Matter (RM) model used here, these problems are circumvented.

The first successful method to interpret a large number of the DIBs was described by Holmlid (2004a), with assignment of more than 60 bands of various widths in a spectral region with a large number of sharp and relatively weak bands. This method used quasi-classical energy level calculations for a special class of doubly excited atomic states. It was assumed that the absorbing entities were ions in a phase called Rydberg Matter (RM), forming doubly excited atoms. The optical transitions were found to be due to electron transfers between excited atoms. This RM material has been studied experimentally in a large number of publications. A later investigation of absorptions in doubly excited atoms and in well-known singly excited He, transferring to doubly excited states, then gave an assignment of a further 63 DIBS, including all the intense and sharp DIBs (Holmlid 2008a). The main characteristics of the DIBs were then clarified. In the present study, practically all DIBs known prior to 2009 are interpreted, using both types of transitions developed in the two earlier studies. Other atoms than He also contribute. A new method of interpreting the very broad DIBs, which were reported in older observational studies, has been added. Such broad bands are of course more difficult to assign due to their low information content. It is shown that the band edge is more important then the band centre for assigning these broad bands, which is not unexpected due to the metallic or even superconductive environment.

Metal atoms form a metallic phase, where the electrons are shared and delocalized in the conduction band in the metal. A normal metal has its conduction electrons in linearly translating states, with zero angular momentum. RM is a metal-like phase where the electrons interact and form a conduction band. However, they still retain orbital angular momentum, each one orbiting around an ion in the material in a Rydberg-like circular orbit. Due to the size of such orbits, the material may be of very low density, but at low values of angular momentum, the material has normal density. The RM phase has an internal structure due to the slow orbiting electrons, which prevent the formation of large continuous volumes by so-called retardation effects (Holmlid 1998). RM thus consists mainly of small planar six-fold symmetric clusters. Stacks of clusters are formed at low enough temperature. Many different experimental and theoretical methods have been used to investigate RM (see e.g. the article "Rydberg Matter" in Wikipedia). RM has been produced mainly from alkali metal atoms and hydrogen atoms in the laboratory. It has not yet been produced in the laboratory from He atoms. A theoretical study by LaViolette et al. (1995) of a condensed excited He phase similar to RM only described the condensation of the He (2 $^3S$) state. The very low level He (2 $^3S$) gives just a few DIBs as shown below, but similar condensed phases for higher excited He states may exist. They may take part in the formation of the DIBs. The main condensed phase involved in the formation of the DIBs is however He(RM).

Doubly excited atomic states with circular electron orbits are used here in the theoretical model. Such levels have been observed within RM in experimental studies with a few different methods, for example in intracavity stimulated emission studies in the IR. A recent publication is found in Holmlid (2007b). Transitions down to principal quantum numbers *n*" = 6, 5 and 4 are observed from slightly higher levels in a $K^+$ ion. This shows that doubly excited states exist within the RM. Similar excited circular levels were observed by radio-frequency stimulated emission spectroscopy of nuclear spin-flips in Holmlid (2009a). The environment for these states was stacks of planar clusters. The mechanism by which the excitation energy is assembled in RM was studied by laser-induced mass spectrometry in Kotarba & Holmlid (2009).

## 2. Rydberg Matter (RM)

The quantum mechanical theory of Rydberg Matter (RM) was developed by Manykin et al. (1980, 1992a, b, 2006). RM is a condensed type of matter formed from long-lived so-called circular Rydberg species, mainly atoms with the outermost electron in an almost circular (high-angular momentum) orbit. An electron is in a Rydberg orbit when it is hydrogenic, thus in the same form as in the hydrogen atom and not appreciably influenced by any other electrons. The atom is then in a Rydberg state. In a circular Rydberg state, the interaction of the circular electron with the inner electrons is minimized. Rydberg states are formed for example by recombination of ions and electrons, a process which is known to be common in interstellar space. Such states have been identified in interstellar space with principal quantum number *n* up to 1000 by long-wavelength radio emission (Sorochenko 1990).

In an ordinary metal, the conduction electrons have zero orbital angular momentum. RM is a generalized metal, with the bonding electrons in Rydberg-like circular orbits with angular momentum different from zero. Quasi-classical calculations of the bonding and electronic properties of RM have shown the material to form small planar clusters (Holmlid 1998). These calculations also proved the direct relation between the electronic excitation level *n* and the interatomic distance in the material. Both the planar cluster forms and the interatomic distances have been verified by rotational spectroscopy (Holmlid 2007a, 2008c). RM is a state of matter comparable to liquid or solid matter and can be formed from most atoms and small molecules. So far, RM in various excitation levels has been formed in the laboratory from Cs, K, $H_2$, $N_2$, H and D. The easiest way to form these types of RM is by thermal desorption of the atoms or molecules from non-metal surfaces like metal oxides and graphite (Kotarba et al. 1995; Holmlid 2002). The RM phase is almost metallic and has a very long radiative lifetime, of the order of hours in the laboratory (Badiei & Holmlid 2006; Holmlid 2007b). In space, its radiative lifetime which is found from RM theory (Manykin et al. 1992b) is close to the assumed lifetime of the universe (Holmlid 2000). RM consists mainly of six-fold symmetric planar clusters with so-called magic numbers (number of atoms) equal to $N = 7$, 19, 37, 61 and 91. RM clusters formed from K have been studied in the laboratory by electronic spectroscopy (stimulated emission in a laser cavity) (Holmlid 2007b), by vibrational spectroscopy (IR laser Raman scattering) (Holmlid 2008b), and by rotational spectroscopy (radio-frequency stimulated emission) (Holmlid 2007a, 2008c). RM formed from H and D is different from other forms of RM in that it does not only form planar clusters, but H(RM) and D(RM) are three-dimensional quantum materials (Holmlid 2008d; Badiei et al. 2009), with special properties like superfluidity (Andersson & Holmlid 2011) and probably also superconductivity. It is expected that He(RM) is similar to H(RM) in these respects. Planar RM clusters form stacks of clusters, in the end giving long filaments of clusters. This was predicted from a calculational study in Badiei and Holmlid (2002c). The recent study on the spin-flips of $^{39}$K nuclei in K(RM) clusters (Holmlid 2009a) shows that such stacks of planar clusters exist. They have also recently been identified directly for H(1) by mass spectrometry (Holmlid 2011). A typical stack of clusters is shown in Fig. 1.

The existence of RM in space has been inferred from several unique spectral and radiative signatures. The so called unidentified infrared bands (UIR, UIB) have been identified as due to RM thermal emission (Holmlid 2000). Such spectra have been studied in detail in optical cavities (Badiei and Holmlid 2005, Holmlid 2004c). This type of emission is the basis for the RM laser (Badiei and Holmlid 2003). Similar spectra are observed from comets (Holmlid 2006a). The polarized scattering of light from comet comae does not seem to be due to ice or carbon particles, but agrees with that expected from RM clusters (Holmlid 2006a). The alkali atmospheres on the Moon and Mercury can be interpreted as due to photo-dissociation of RM clusters, with the clusters at high altitudes acting as a reservoir of alkali in the atmospheres (Holmlid 2006b). Several features of the atmosphere of Mercury are explained in this study, like its large electrical conductance. At the mesopause in the atmosphere of Earth, alkali metal, especially Na, accumulates and form huge layers. If these layers are in the form of RM, they can have helped to give the homochirality of living organisms by forming circularly polarized light (Holmlid 2009b). Maser emission is shown to be due to optical amplification caused by accidental agreement of singular rotational lines from small molecules with rotational lines of RM clusters formed from hydrogen (Holmlid 2006c). This explains how masers with extremely high energy can be formed, since they are unlikely to be supported by any pumping mechanism in small molecules. The Faraday rotation observed at radio frequencies in intergalactic space can be explained by a low density of RM (Badiei and Holmlid 2002b).

## 3. Calculational method

The same method as developed and proved to work in Holmlid (2004a, 2008a) is used here. This means that optical transitions in a phase of He(RM) are calculated, using known singly excited states of He atoms and doubly excited states of He atoms (which are less well-known and will occupy us somewhat below). The doubly excited states are not free atoms but they are part of the planar RM clusters. The clusters will in general not be free but will exist within stacks of clusters, as described above. The planar environment will directly influence the electronic states and their orientation. All quantum numbers used to describe free atoms are not good quantum numbers in this case, and the principal quantum number $n$ is not always a valid quantum number. This is so since the potential energy does not vary as $r^{-1}$ due to the external field and due to the planar charge distribution of the inner electron in the doubly excited states. The orbital and spin angular momenta of the electrons are however good quantum numbers, if the relatively small effects due to their coupling in the RM clusters are neglected in first approximation. The typical cases of a free RM cluster and a cluster within a cluster stack are shown in Fig. 2, with the quantities shown defined in the figure caption. In most cases the atomic electronic state can be quite well described by the orbital angular momenta, with the spins added afterwards as in the simplest free-atom case. Thus, a classification similar to L-S coupling is used here for the doubly excited atomic states.

Other coupling schemes are however possible but the complexities of a full cluster angular momentum coupling scheme (as seen in Fig. 2) do not seem to be required for a reasonable description. However, some departures of the order of 10 $cm^{-1}$ (approximately 0.1% of the energy of the state) may be due to such effects. This level of accuracy of the calculations is overall extremely satisfactory.

The calculational method will be summarized. Since a time-independent solution is sought, the energy levels should be independent of the special position of the electrons during the motion. The neutral atomic state is thus modelled as a core ion with charge +2, with an inner electron in a circular orbit in the field from the ion, and an outer electron in a circular orbit in the same plane as the inner electron. Similar doubly excited states are found in several experimental studies described above (Holmlid 2009a, 2007b). No quantum-mechanical description of this type of doubly-excited atomic state is yet possible, and we thus have to rely on a quasi-classical approach with a quantum defect description. The quantum defect describes the small effects due to the many-body effects in the interaction between the electron and core ion with its electron(s), which is often referred to as core penetration of the outermost electron (Gallagher 1994; Rau & Inokuti 1997). Here, the main factor taken into account by the quantum defects is the electron spin interactions giving triplet states.

The shielding of the core ion due to the inner electron is calculated by averaging the interaction potential energy during one period of the electronic motion. The inner electron in the atom is in its Bohr orbit with principal quantum number $n_3$ and with $l$ =

$n_3$, thus in a classical circular orbit. The integral for the potential energy of the outer electron is given by

$$V_4 = -\int_{-\pi}^{\pi} \frac{e^2}{8\pi^2\varepsilon_0}\left(\frac{2}{r_4} - \frac{1}{(r_4^2 + r_3^2 - 2r_3r_4\cos\theta)^{1/2}}\right)d\theta \qquad (1)$$

where $r_4$ is the orbit radius for the outer electron and $r_3$ the same for the inner electron. In effect, the outer electron is pushed outwards relative to the case with a spherically distributed inner electron charge, as shown by Holmlid (2004a), Table 2.

This integral in Eq. (1) is evaluated recursively, with the initial guess of $r_4$ as $n_4^2 a_0$ where $a_0$ is the Bohr radius. When a positive quantum defect is included, $n_4$ is decreased by this amount. The orbital angular momentum for the outer electron is calculated as $pr_4 = n_4\hbar$. This gives the kinetic energy and the total energy for the outer electron. The derivative of the potential energy in Eq. (1) is used to determine the stable circular orbit with the correctly quantized orbital angular momentum. The new value of $r_4$ is used as a better value of $r_4$ for a renewed calculation of the potential energy until a stable value of $r_4$ is found. This means that the calculation of the potential energy $V_4$ is for a fixed quantum number $n_4$ for the outer electron, and that the orbit is circular for the electron with $n_4 = l_4$. It should be noted that the potential for the outer electron in the present case is not easily calculated in an exact classical way in the form of a closed orbit since the force is non-central. This implies that the principal quantum numbers *n* are not good quantum numbers, but *n* is used here instead of *l* for simplicity. To simplify the description of the states, the inner electron number value is designated as *n*" and the outer (valence) electron number value is *n*'.

## 4. Results and discussion

The results for all the transitions calculated are given in a table which can be found in the additional material on the web, here indicated as contained in the Appendix. The list of DIBs used is based on the results by Galazutdinov et al. (2000) (GRMW). Results from the UV and NIR ranges taken from Jenniskens and Désert (1994) are discussed below. The equivalent widths and the FWHM of the bands in the table on the web are from Jenniskens and Désert (1994) to simplify the analysis of the results. Upper and lower states of the transitions, as well as the specific atom giving the transition (if applicable) and the publication (Holmlid 2004a, 2008a) where the band was first assigned as given here are included. Below, these results are grouped together from various aspects and discussed. Observational results on the shape of the bands and correlations with other lines mainly concern the most intense DIBs, and such results were discussed at length in Holmlid (2008a) which primarily treated the sharp and intense DIBs. The description here is mainly given in nm and in $cm^{-1}$ which are the normal spectroscopic quantities used, while all tables also include data in Å to simplify comparisons with previous studies.

### 4.1. Co-planar doubly excited atomic states

Most sharp and intense DIBs were interpreted in Holmlid (2008a). A few strong DIBs in the GRMW range and outside the GRMW range were not interpreted in Holmlid (2008a). Inclusion of such transitions (and others) means that a few more co-planar doubly excited atomic states have been added to the ones used in Holmlid (2008a). Table 1 contains the relevant information. The symbols used to describe these states are of the form $n_4/\ n_3$, where $n_4$ is the quantum number for the outer electron as clearly indicated in the tables. (The spin multiplicity is appended to the upper left corner of this symbol, as normally done). In all, 21 quantum defects different from zero are used. The quantum defects are only allowed to have relatively small values, with a maximum value of 0.05 in Table 1. This is considerably smaller than typical quantum defects for p electrons for alkali atoms, but comparable to those for d electrons in light alkali atoms like Li and Na (Gallagher 1994). Compared to singly excited Rydberg states of He, the values found are similar to those for P and D states (Gallagher 1994). Thus the core penetration of the electrons is small as expected in the present case of circular Rydberg orbits. The main effect giving quantum defects different from zero here is the interaction of the spins of the two electrons to form triplet states. This agrees with the fact that the singlet states often have close to zero quantum defects, as seen in Table 1. In the case of singly excited He Rydberg states, the triplet states also have larger quantum defects (Gallagher 1994) in agreement with this. For singly excited He Rydberg states, it is known that a transition to mixed singlet and triplet states occurs at large angular momentum, with variable admixture of the other multiplicity. The small quantum defects give small changes in the energy relative to zero defect, and negative quantum defects change the levels considerably less than positive defects. Thus, the calculated levels are only modified slightly by the quantum defects used. Since 21 different defects are used to give the energy levels and 260 transitions are calculated from them with small errors relative to the observed DIBs, this set of parameters seems quite reliable.

For reference, a new set of energy values for the doubly excited states with zero quantum defect is given in Table 2. The changes relative to the table published in Holmlid (2004a) are small, but the values are now based on the Rydberg constant for He.

## 4.2. Intense DIBs involving doubly excited states

All intense DIBs from the GRMW study are interpreted in Table 3. All bands indicated in that study as intense with a transmission <0.95, or with an indication as broad (br) are included. Several different types of transitions can be observed in this table, and the two main types will be described here. (The transitions to Rydberg levels will be discussed below). One of the strongest band which is at 15116 $cm^{-1}$ is identified as $^{1}3/1 \leftarrow {}^{1}2/1$, thus a spin-allowed singlet-singlet transition. The probable mechanism for this transition is shown in panel A in Fig. 3. It is an excitation of the outermost electron in an atom, where in fact the inner electron is in its lowest state. The outermost electron is excited from $n'= 2$ to $n' = 3$. (A similar interatomic transition, as below, is also possible). The related triplet transition $^{3}3/1 \leftarrow {}^{3}2/1$ is observed at 15825 $cm^{-1}$ but is of lower intensity and not included in Table 3. (Both these transitions are allowed by spin and angular momentum selection rules). This type of transition within an atom is quite common among the intense bands in Table 3. However, also examples like $^{1}3/4 \leftarrow {}^{1}2/2$ at 15680 $cm^{-1}$ are identified. In this case, the electron is excited from one atom to an ion, as shown in panel B in Fig. 3. Such a transition is of the form

$$\mathrm{He}(2/2) + \mathrm{He}^{+}(-/3) + h\nu \rightarrow \mathrm{He}^{+}(-/2) + \mathrm{He}(4/3) \quad (2)$$

where - indicates an electron missing in the He atom. This may be thought of as an electron jump from one $He^{+}$ ion to another in the absorption process.

Another type of transition is also found in Table 3. It involves a singly excited He state as the initial state and a doubly excited state as the final state, for example $^{3}2/4 \leftarrow 2\ ^{1}P$ at 14679 $cm^{-1}$. Such a process is most likely an electron transfer similar to that in panel B in Fig. 3, but from a He atom not (yet) embedded in the RM cluster field and thus retaining the free-atom electronic levels. The He atoms in contact with the RM phase may be influenced by the field from the RM, and thus the free He atom energies may not be exactly applicable. However, the level of agreement found is certainly sufficient for accurate assignment of the bands. In general, the agreement is quite satisfactory with an average absolute error of 4.3 $cm^{-1}$ for the 27 bands interpreted in Table 3, slightly larger than the half-width of these sharp bands.

## 4.3. Broad DIBs

All broad and intense bands not marked uncertain (but three) are interpreted in Table 4. This table is based on the DIB survey by Jenniskens and Désert (1994) and all bands with an equivalent width larger than 0.1 Å or a real halfwidth (FWHM) larger than 10 $cm^{-1}$ are included. Some of the bands in Table 3 also exist in Table 4, however the many shallow broad bands detected in older studies dominate in Table 4.

In Table 4 many assignments given are similar to the ones for the sharp and intense bands in Table 3. However, the centre of each observed band is slightly below the calculated nominal transition. The mechanism of the excitation is probably in some way related to the broad conduction band in the RM material. If the conduction band is directly involved, the transitions are not expected to take place within just one atom, as a few of the intense and sharp bands discussed above were concluded to do. It can be observed directly in the table that the broad bands are not related to atomic excitations with the lower electron unchanged, but always involve an electron transfer from one atom to another. For example, the intense band at 20479 $cm^{-1}$ is interpreted as being due to a transition 5/4 ← $^1$2/2 calculated to be at 20525 $cm^{-1}$. The proposed mechanism in Fig. 4A shows an electron transfer from an atom (doubly excited) to an ion. The transfer energy is lower than calculated nominally, which may be caused by a variation of the Fermi level in the condensed metallic RM phase in Fig. 4A. In such a case, the width of the transition does not have to be directly dependent on the conduction band electrons, but the width of the transition may be due just to a distribution of the Fermi level in the material. Another possibility is an excitation to a low level in the conduction band and a simultaneous deexcitation of an electron from a higher level in the conduction band, thus a two-electron process as shown in Fig. 4B. Such a mechanism may be possible with strongly coupled electrons as in a superconductor. In a few cases in Table 4, the energy difference is found to have the contrary sign (positive), but no conclusion on the origin of this effect has been reached. In principle, both signs are possible depending on the details of the RM environment like the variation of the Fermi level with position.

Another type of broad band also exists in Table 4. This is a transfer of an electron from a doubly excited atom to a Rydberg level around an ion, of the form /7 /2 ← $^3$2/2 calculated to give 16231 $cm^{-1}$ but assigned to the band with its center at 16184 $cm^{-1}$ with a width of 60 $cm^{-1}$. This transfer is shown in Fig. 5, panel A. In this case, the final electron can only be in a state in an ion in the RM phase since the value *n*" = 7 is odd. With an even value of *n*" as for 20724 $cm^{-1}$ with *n*" = 10, the transition could be to a state in an atom, thus at $n' = 5$. No case with a transfer from a singly excited He atom to such an ionic state is detected.

Another type of broad band was discussed already in Holmlid (2004a) and gives the so called band-heads indicated in Table 4 at 16633 and 15305 $cm^{-1}$. A few other band-heads are overlapped by other bands. The band-head at 16633 $cm^{-1}$ is the lowest energy transition for the series of transitions 11/1 ← /5 /1 with /5 meaning an ion with the electron at $n$" = 5, and $n$" = 1 being the number changing in the series. This mechanism with $n$" = 1 is shown in Fig. 5B, and the process can be written as

$$He^{+}(-/5) + He^{+}(-/1) + h\nu \rightarrow He^{2+}(-/-) + He(11/1) \quad (3).$$

The higher transitions in the series have the form

$$He^{+}(-/5) + He^{+}(-/n) + h\nu \rightarrow He^{2+}(-/-) + He(11/n) \quad (4)$$

with increasing $n$ up to twice the value of $n_4 = 11$. Higher band-heads like 16/1 ← /5 /1 exist but at lower intensity and as sharper bands. The lowest band-heads 6/1 and 5/1 probably exist close to 14407 and 13207 $cm^{-1}$ but the mismatch is larger than the bandwidths, probably since the quantum defects for those low (triplet and singlet) states are not negligible (not included in Table 1)..

In a few cases in Table 4, a transfer of the same type as mainly discussed in Holmlid (2004a) and just above is observed, giving broad bands which are not band-heads. One example is the band centered at 15496 $cm^{-1}$. It is remarkable that these bands are all due to transitions from $n$" = 4, not from $n$" = 5 which is the most common value for this type of DIB. Since $n$" = 4 is at the same energy level as $n' = 2$, a normal Rydberg level even in a H atom, this energy easily leaks away via the H atoms believed to surround the He parts of the interstellar clouds. Thus special conditions are probably required for the support of the *n*” = 4 electron state, possibly a continuous He(RM) medium (without H(RM)) which will support the transitions by electron coupling through the conduction band, giving the broad bands observed in this case. For the more normal value *n*” = 5, a well-formed He(RM) is not required since H atoms cannot remove this energy rapidly, and the resulting DIBs are sharp with no involvement of the conduction band. Thus, they do not appear in Table 4 but in the other tables with sharp transitions.

## 4.4. Transitions to RM and Rydberg states

The model of Rydberg atoms and RM as the material in which the dominating doubly excited states are embedded gives assignments of most DIBs accurately as described above. It is expected from this model that transitions to the Rydberg limit or most often to the lower edge of the conduction band in RM (at approximately 50 $cm^{-1}$ below the ionization limit) will exist for the doubly excited levels given in Table 1. As already shown in Holmlid (2008a), such transitions are found. Many of the levels in Table 1 will give transitions far outside the normal DIB range and such bands have not been detected. However, a number of bands are observed as shown in Table 5. The existence of these bands is strong support for the accuracy of the results found. It is especially remarkable that accurate values are predicted for the four singular DIBs observed in the range 803 – 850 nm at 12438, 12303, 12069, and 11846 $cm^{-1}$. Most of the levels which give rise to these DIBs are calculated using quite small values of the quantum defects, which means that the theory is accurate.

Other similar transitions are also expected. The most common level for many weak transitions is *n*" = 5. In Table 3, the transition at 17523 $cm^{-1}$ agrees very well with that expected for an RM transition, with a transition close to 17555-50 $cm^{-1}$. This band is observed to be both broad and intense. Similar transitions from *n*" = 4 and 6 either fall outside the DIBs range, or in a less sensitive range of the DIBs as in the case of *n*" = 6 at 12191 $cm^{-1}$. Several singly excited He states are noted above to take part in transitions involving electron transfer. Such states are also expected to give similar transfer transitions to the RM conduction band or to the Rydberg limit. Three such transitions observed within the range studied are included in Table 5 in the bottom lines. It is notable that these transitions are matched by theory more accurately than those from the truly embedded doubly excited states in the top of Table 5. Thus, it seems that the transition to the Rydberg limit at least for the two sharp lines at 15074 and 15072 $cm^{-1}$ is to the atomic Rydberg limit, not to the RM limit.

That sometimes a broad band and sometimes a sharp line is observed for the Rydberg and RM transitions may partly due to background effects due to spectral congestion in the measurements. Close to the red end of the DIBs range, only sharp peaks seem to be observed in agreement with this suggestion.

Several sharp almost equidistant bands exist in the range 17100-17350 $cm^{-1}$. Several such bands were identified in Holmlid (2008a) as being due to transitions to the RM limit as

$$He^{+}(-/5) + h\nu \rightarrow He^{+}(-/n") \qquad (6)$$

In Table 6, several more bands are included, giving observed values of final *n"* in the whole range 31-47. This is within 500 $cm^{-1}$ from the absolute upper limit of the RM conduction band and thus far below the normal conduction band in RM. This series of bands shows directly that Rydberg states are involved, and since odd values of *n*" exist it is definitely proved that the transitions are due to the outermost electron in a Rydberg ion (only even values might mean an electron in a Rydberg atom). The cut-off to higher values of *n"* in this case can be interpreted as a limit in the distance between the ions in the surrounding RM phase, i.e. that the structure has this maximum size and no larger values of *n"* can be accommodated within the RM structure. This ion-ion distance in RM at *n"* = 47 is 0.34 μm. The corresponding series from *n*" = 4 and 6 should be found at 26943 – 27240 and 11704 – 12001 $cm^{-1}$. These regions have probably not been studied with the same sensitivity as the bands from *n*" = 5. Since these even values of *n*" can transfer their energy to H Rydberg states, they are probably less populated (see further below). Stimulated emission experiments with RM at high temperature show a maximum excitation level of *n*" = 40-80 (Holmlid 2004c). An analysis of the unidentified infrared bands (UIR, UIB) as emanating from transitions of the inner electrons in RM gives an excitation level of at least *n*" = 40 (Holmlid 2000). Most UIR bands are observed at *n*" = 40-80, but the UIR emitting clouds are probably warmer than the ones giving the DIBs in absorption. The good agreement in Table 6 shows clearly that inner electron levels are an important part of the DIB spectra.

One intriguing effect exists in this RM series of bands, and that is the very strong band at 17295 $cm^{-1}$, in fact one of the strongest DIBs i.e. the original 578.0 nm DIB. This band seems to belong to the RM series, and behaves differently in the ISM clouds than the closely located and equally strong band at 579.7 nm (Galazutdinov et al. 1998). This different behaviour was explained as being due to the different origins of the bands, with 579.7 nm originating in absorption in He (2 $^3$P) (Holmlid 2008a). This explanation does not however cope with the very high intensity of the absorption at 578.0 nm, i.e. the reason why the band at *n*" = 41 becomes so strong relative to the other bands in Table 6, and this is developed in the following description. In Tables 3 and 4, another process is held mainly responsible for the 578.0 nm DIB, adding a process starting from He (2 $^3$P) as for the 579.7 nm DIB. The upper state matches a state named $^0$3/2, which has a small quantum defect and is included to match a few other lines. It may be a level in 3/2 which is separated by angular momentum (J splitting or similar), but the ordering of the levels is unknown and probably not straight-forward judging from other He Rydberg states (Gallagher 1994). It is likely that this transition is the basic resonance, but the matching with the transition *n*" = 41 ← 5 gives added intensity due to the resonance to the upper RM levels. Another factor which may increase the intensity of the 578.0 nm DIB (and also of the 579.7 nm DIB) is the proximity of the 3 $^1$P state of the He atom to the RM band. This level is 17123 cm$^{-1}$ above the initial state 2 $^3$P, just where the RM series stops towards lower *n*" in Table 6. Thus, it appears that several factors due to the RM environment increase the intensity of the DIBs in this spectral region. In fact, both these DIBs are found on top of broad bands (Jenniskens and Désert 1994) possibly due to such other optical effects. The main factor which promotes 578.0 over 579.7 is probably the accidental matching of 578.0 with *n*" = 41 ← 5, while 579.7 falls in the middle between the RM series peaks. Thus, environmental factors like the density and general excitation level of the RM phase may influence the relative intensities of these two bands. This may explain the observations by Galazutdinov et al. (1998).

## 4.5. Weak bands

In the main table in the added material to this publication, most DIBs interpreted are relatively weak and narrow. This type of DIB was described in Holmlid (2004a) and is due to transitions similar to the one shown in Fig. 5B. An example is provided by the band at 16482 $cm^{-1}$. It may be described as 10/7 ← /5 /7, where /5 is a Rydberg ion with the electron in *n*” = 5. Initially, two $He^+$ (or metal) ions exist in the RM phase, one with the electron in *n*” = 5 and the other with *n*” = 7. During absorption of a photon, the electron from *n*” = 5 is transferred to a state *n*’ = 10 around the *n*”= 7 electron giving one doubly excited atom and one doubly charged ion. This is similar to the process in Fig. 5B but with other quantum numbers. Such bands are most easily calculated for large quantum numbers where the quantum defect is negligible. It seems now that most weak DIBs involve large quantum numbers and thus negligible quantum defects, in contrast to the view held in Holmlid (2004a, 2008a). This means that it is in principle possible to assign all weak DIBs. However, a few different assignments are in general possible for each band, and a list of possible assignments is not very informative. Several weak bands are in the main table shown to agree accurately with transitions from singly excited Li and Na atoms to doubly-excited states. This is the same type of transition as discussed above for singly excited He atoms. Many of the remaining weak bands may be due to similar transitions involving other singly excited atoms like C and O. However, the assignments of such transitions are quite uncertain if not a large number of bands agree well. Further observational studies would be of interest in this respect.

## 4.6. Energy level diagrams

To supplement the discussions above, the transitions only involving co-planar doubly excited states in Tables 3, 4 and 5 are included in energy diagrams for such states in Figs. 6 - 8. From these figures, it is immediately clear that transitions between states in the upper half of the diagrams will fall outside the visible range, and such transitions are thus not observed. The limit of the DIBs towards the IR is approximately at 11 000 $cm^{-1}$, and this limit almost prevents levels above 2/4 to be the lower state in a transition to higher states. Many other factors may also decrease the probability of such transitions to high states, like angular momentum conservation and insufficient space for the large orbits. In the case of direct transitions as in many cases in Fig. 6, the small overlap between low and high states will be highly restrictive.

A comparison can be made between Figs. 6 and 7, showing the intense bands from Galazutdinov et al. (2000) in Fig. 6 and the broad bands from Jenniskens & Désert (1994) in Fig. 7. Some of the transitions in these two figures are the same, as can also be seen in the tables. The sharp bands are observed to go to intermediate states, almost all below 2/5, while most of the broad bands go to states higher than 2/5. This clear difference in behaviour is understandable. The transitions giving the broad bands usually involve electron transfer via the conduction band. Thus, the state of the inner electron in the transfer is not important and all transitions are in principle possible. The fraction of transitions that may be intra-atomic is much smaller for the broad bands than for the intense bands, in agreement with this. For the sharp and intense bands, many of them can be intra-atomic and thus transitions to high levels with large values of *n*” are not possible since the absorbing low states have low values of *n*”. It was shown in Holmlid (2008a) that weak selection rules exist for the intense and sharp DIBs, and some limitations in the short-distance electron transfer possible for the intense bands may exist. The type of atoms involved (He or metal atoms) may also influence the electron transfers, as discussed further below.

## 4.7. Accuracy of the calculations

The assignments of the intense and narrow bands in Holmlid (2008a) and Table 3 are mainly based on known atomic levels and on the fitting of the band positions with the help of a small number of quantum defects, as described in detail in Holmlid (2008a). Thus, the agreement of the intense bands with calculated values is mainly a measure of the accuracy of the assignments of the various levels. An agreement of the order of a few $cm^{-1}$ is considered to show that a useful description of the levels has been made. However, in some cases in Table 3 the differences between observed values and the calculated ones are larger than 10 $cm^{-1}$, which is significant. In some cases this is probably caused by further splitting of some of the doubly excited states in sublevels due to angular momentum, as included tentatively in the case of the levels named $^{0}3/2$ in Table 1. Due to the large angular momentum in electron orbiting motion and the co-planar atomic structure, cases with different properties due to different signs of the angular momentum for the inner electron in $He^+$ are certainly likely. This would give total $L = 1$ or 5 in the case mentioned, with further adding of $S$ to form a $J$ vector, if this is the process followed for the angular momenta in these atoms. If this will give further selection rules is not known.

In the case of weak bands which depend only on calculated levels for doubly excited atoms, the accuracy aimed at is also a few $cm^{-1}$. The overall agreement with the DIBs is of this size, but in many cases the assignments are not known from other sources or can be inferred from intensity or band half-widths. Thus, the matching may for example be skewed in a series of bands which stretches over several hundred wave-numbers. The general impression after matching the bands is that the number of weak bands which have not been detected is still rather large. In fact, spectral congestion seems to exist for example close to the band-heads and other similar regions. This makes the background level change due to the unresolved features, and this may be one reason for the many broad DIBs recorded. Thus, weak DIBs are often observed as a result of spurious overlap of several bands from different sources. Unfortunately, it is not possible to model the total spectrum by computer since basic spectral features like intensity variations are not known, and such information will only be available after extensive laboratory studies and further theoretical work.

## 4.8. Atomic forms

Since many of the intense DIBs involve singly excited He states, the dominating importance of He for the creation of the DIBs is clear relative to Li, Na and other metlas. The involvement of the low doubly-excited co-planar states like $^{3}2/2$ in almost all DIB transitions indicates further that He is the optically most important species in the RM condensed matter, which is the basis for the interstellar clouds where the DIBs are observed. It might be expected that H atoms as the most abundant species would also have a role, but no evidence for this exists. Not even in the series of transitions to high RM levels do H atoms take part, since the transitions only involve the quantum number *n*” for the inner electron in He or some other two-electron atoms.

The weak DIBs are almost all derived from electrons in *n*” = 5, which cannot be a level in anything but a two-electron atom like He: such levels do not exist in H atoms or around a $H^+$ nucleus. If such two-electron atoms can be heavier atoms like Ca or other metal atoms is not immediately clear. The orbiting distance of the *n*” = 5 electron is 0.66 nm, which is relatively large. For comparison, the ionization energy of such an electron is 2.2 eV, thus considerably smaller than most valence electrons in an atom. This electron will thus be at a larger distance from the nucleus than the other electrons in the atom. The ionic radius of $Ca^{2+}$ is normally considered to be < 0.1 nm, while $Mg^{2+}$ has a radius < 0.07 nm. Most other metal ions like $Fe^{2+}$ are smaller. Thus, for many metal ions the state *n*” = 5 is possible, which means that they can be involved in the formation of weak DIBs. In fact, they may not only be one of the initial atoms but also the final doubly excited atom, for example in a state 6/6. However, it is not possible that metal atoms can form states like 5/1 or 10/1, since the inner electron indicated cannot exist due to the ion core of the metal. This may be one reason why transitions to band-heads like 6/1 ← /5 /1 are clearly separated in intensity from transitions like 10/7 ← /5 /7. In the first case shown, He is at least one of the two nuclei in the process, while in the second process, almost any atom (but H) may take part as both atoms.

There are a number of DIBs given in the GRMW based table which are not possible to assign within the model of doubly excited atoms, mainly formed by He. Several of these match transitions with singly excited Li and Na states instead of the commonly observed singly excited He states. This means that an electron transfer takes place from a singly excited metal atom forming a doubly excited state. Many more overlaps exist but only the very clear cases where no other assignment is possible have been matched to transitions involving Li and Na in the table. These DIBs are rather weak, but the good match indicates that alkali atoms take part in the formation of the RM, as expected.

Other types of transitions also exist in the tables, involving Rydberg or RM levels as the final states. One of the transitions ending in Rydberg or RM states in Table 5 involves just one Rydberg electron. This is the band at 17523 $cm^{-1}$, corresponding to an excitation Ry $\leftarrow$ *n* " = 5. This transition can take place in any ion that can support a stable *n*" = 5 level, thus many common metals or He as described above. Thus to generalize, the DIBs observed involve both He for the strongest transitions and metal atoms for the weak transitions.

## 4.9. DIBs outside the normal spectral range

Since the different types of DIBs and their origins now are known, it is possible to discuss if and in which spectral range bands of similar origin can be observed. The intense bands are due to transitions involving doubly excited co-planar atoms and also singly excited He atoms. The spectral range for the DIBs is given approximately by the largest energies below the ionization limit in this system of He atoms. This means that no DIBs due to doubly excited species alone can be expected above 28075 $cm^{-1}$ or at wavelengths < 350 nm. The only possibility would be transitions from the He singly excited state 2 $^3$S at 38455 cm$^{-1}$ below the ionization limit. (That the ground state of He at 198311 cm$^{-1}$ below the ionization would be involved in similar transitions seems unlikely). The 2 $^3$S state may lead to bands similar to DIBs in a range below 33000 cm$^{-1}$, extending over the normal DIB range as well. However, just one DIB is interpreted as due to this state namely 10379 cm$^{-1}$ at the lower (long wavelength) limit of the DIB range. The other possible intense transitions from this level to doubly excited states are given in Table 7 as an aid in their future identification.

If we instead of the UV range discuss the IR range, it is apparent that many DIBs may be observed there. For example, all transitions between doubly excited states of the general type x/y ← 3/*n* will be observed below 12000 cm$^{-1}$, or at wavelengths > 830 nm. Also higher singly excited He atoms may be involved. The number of transitions will however be large and thus the peak intensity low, and the number of atoms in such high states above their respective ground states may be low if the cloud temperature is not high enough. A good reason to believe that such transitions will be observed is anyhow the high excitation level in the RM observed from the DIBs, with the general level of electrons at *n*" = 5 at 17556 cm$^{-1}$ below the ionization limit. This may carry enough energy over to the 3/*n* states. Note also that many weak DIBs are due to transfer from states with *n*" = 10 and higher, which are at high energy only 400 cm$^{-1}$ below the ionization limit. This indicates that the excitation may be high enough to observe bands due to transitions x/y ← 3/*n*.

Finally, it is of interest to discuss the possibility of finding DIBs similar to the weak and numerous bands in other spectral ranges. The central point for such bands is the general excitation level, in the present case at *n*” = 5 giving the DIBs gravity point at 600 nm. Higher *n*” levels may be less likely, so we will first investigate the case of *n*” = 4. This corresponds to 27000 $cm^{-1}$ or an expected center of 370 nm. Investigations in the UV have not been entirely successful, but it is possible that the weak bands will be difficult to resolve and that the spectrum will appear congested. Since *n*” = 4 is at the same level as *n*’= 2 which is a state also for H atoms, it is likely that energy cannot accumulate in this excitation level within the He(RM) parts of the clouds (maybe physically small). Thus, it is reasonable to instead consider the *n*” = 3 excitation at 48000 $cm^{-1}$ which will not easily be dissipated into the H(RM) part. This corresponds to a wavelength of 210 nm, slightly shorter than the UV extinction bump at approximately 220 nm. It is suggested that lines analogous to the weak DIBs but starting from the *n*” = 3 excitation level may be resolved in the extinction bump. It should be remembered that it is only the bands corresponding to the numerous weak DIBs which can be observed there, since the intense DIBs have another origin related to the doubly excited states, as discussed above. These intense DIBs cannot be translated into this UV range.

## 4.10. Environment for the DIB carriers

The transitions assigned to explain the DIBs are quite unique. They are obtained by a theoretical method calculating the energy levels of the doubly excited states from first principles, in this case in an electrostatic model with angular momentum quantization for the electrons. The electrostatic model is based on extensive experimental information about the RM condensed phase. Thus it is feasible to discuss the environment for the doubly excited species and the properties of this environment which makes the transitions possible.

The co-planar form of the doubly excited states, with both electrons moving in the same plane, agrees well with the know form of RM clusters. Certainly, the spectroscopic and mass spectrometric studies giving evidence for the planar cluster forms (see e.g. Holmlid 2008c; Wang and Holmlid 2002) have mainly employed RM clusters of K and H atoms and $H_2$ molecules. Thus, it is not *a priori* certain that this form will exist also for He(RM), especially since it is known that H(RM) in its lowest excitation level $n = 1$, thus indicated H(1), is indeed not only planar but also forms three-dimensional clusters (Holmlid 2008d). These clusters are shell structures, and the bonding orbitals are donut-formed, still with some directivity. It appears on the other hand likely that as soon as at least one electron exists on the ion core with the conduction electrons delocalized, the preferred form will be planar. Certainly, the states observed from the DIBs have high orbital angular momenta in general and the shape of the clusters containing them will thus be planar. This means that the clusters also form stacks of clusters (Badiei and Holmlid 2002c; Holmlid 2011) at low enough temperature. This temperature will vary with the cluster excitation level since the bonds strengths also between the clusters vary with the excitation level. See the theoretical section and Fig. 1 for a view of the cluster stacks.

From the description given above of the mechanisms of the various transitions, it is clear that just a few of them (certainly some of the most intense) are due to interatomic transitions, with one electron being excited as shown in Fig. 3A. In Table 3, one third of the transitions are of this type including transitions to RM levels, but several intense DIBs are not. In Table 4, only 13% of the bands are of this type. For most transitions, an electron transfer takes place from one atom (ion) to another. The involvement of the conduction band in giving the broad bands in Table 4 also indicates that the processes are complex. Electron transfer processes are quite well-known for example in biological systems and the solid state, but they usually require that a stationary state (quantum mechanically defined) exists, and that the energy required for the absorption process is equal to the transition energy to the state where the electron can be transported. Here, the processes appear to be a transfer of an electron from one ion to another, possibly by tunnelling between the ions or by formation of a transient volume with no electron density in the electron cloud between the atoms which will simplify the transfer. The alternative is a two-electron process where one electron is excited in one atom and another electron is simultaneously deexcited in another atom. Such processes which are separated by some distance probably require that the electrons are strongly coupled in some sense. Due to the strong magnetic fields from the electrons with large angular momenta, special features may exist. For example, nuclear spin-flips were studied in K(RM) giving the magnetic field at the nucleus with high precision (Holmlid 2009a). This process depends on the coupling between spins in several clusters in a stack, thus on a long-distance contact in the RM. Due to the large number of broad bands which indicate a coupling through the conduction band, it is suggested that two-electron processes with strongly coupled electrons at some distance dominate in the DIBs. This may mean that the He(RM) phase is a superconductor, at least over short distances in the clusters and clusters stacks

From an observational point of view, it is interesting to discuss the nature of the ISM and the clouds obscuring the stars used for DIB observations. The information from the DIBs shows that the clouds contain large amounts of metallic or conductive clusters, often forming filaments of stacked clusters. They are “dark” (meaning non-absorbing and not observable) due to the lack of moving electric or magnetic dipoles which can interact with the radiation fields, but they will nevertheless be able to scatter or reflect radiation from the stars due to their metallic properties. The diameter of a typical $He_{19}(40)$ cluster, a 19-atom cluster at excitation level 40, is 1000 nm, thus large enough to scatter visible light. The stacks of clusters will have dimensions a few orders of magnitude larger than this and will scatter light. In this phase, also other atoms like hydrogen and metal atoms are easily incorporated, but due to the resulting rotating dipole (center of charge displaced from center of mass) such clusters will emit radiation at radio wavelengths. It is thus suggested that the reddening of the light in the DIB clouds is due to the large RM clusters and cluster stacks.

## 4.11. Intensities and densities

The relative intensities for the different initial states of the DIB transitions can be found by summing the equivalent widths of the bands observed using the data in the DIBs survey by Jenniskens and Désert (1994). The results are shown in Table 8, both for doubly excited atomic states and for singly excited He states, using the sets of bands in Tables 3 (intense) and 4 (broad). By comparing for example the related triplet and singlet states for the doubly excited states, summed or separately, it is observed that the higher singlet states are equally or more populated relative to the lower triplet states. Thus the RM giving these absorptions is not at high temperature but in an inverted state.

This is in agreement with the basic inverted property of the RM phase giving stimulated emission (Holmlid 2007b). The conclusion from the singly excited state intensities is less certain since these values are more random and not so easy to compare due to lack of suitable related bands. However, it appears that the distribution of these states is more thermal, since the lower triplet states in the sum are more abundant. This is expected since the singly excited He atoms are not part of the inverted RM but only attached to the material, probably on the surface of the stacks of clusters. The graininess of the data for the singly excited states is however large. The reason for the 2 $^3$P being the most frequent initial state may simply be that this is the highest angular momentum state, with L+S = 2. This may give easier transfer to the high angular momentum doubly-excited states. The overall intensity of the DIBs, both intense and broad is mainly due to low doubly-excited states, with only roughly 30% of the intensity due to absorption in singly excited He atoms. As an illustration of this, the bands with the strongest absorption are shown in Fig. 9. Assignments due to absorption in doubly excited states are indicated with bold and italic font, and absorption in singly excited He atoms with normal font. The band at 578.0 nm is boxed, indicating a mixed RM origin, as described above in section 4.4.

The energies of doubly excited atomic states like the He states of importance here can not yet be calculated quantum mechanically. This is not a special feature of the RM or of the DIB transitions, but a general observation. As discussed in Holmlid (2004a), the precision of quantum mechanical calculations for systems of a complexity similar to RM clusters is at least two orders of magnitude (thus a factor of 100) *worse* than for the quasi-classical calculations used here. It is of course necessary to assign the transitions first, before any extensive calculations of the rates of transition, thus of the band intensities, can be attempted: otherwise the calculations cannot be done since the states involved and the transition processes are not known. It is also necessary to assign all possible DIBs before intensity calculations are investigated, since overlaps of different DIBs are common.

No accurate method is known at present to calculate the rates of the various transitions giving the DIBs. However, if the densities of the DIB carriers are unknown since they cannot be measured because the DIB carriers are unknown, there is no need to develop a theory to calculate transition intensities. The transition intensities would not be useful to calculate the DIB intensities when the densities are not known. It may still be interesting to discuss the prospects of theoretical advancements in this direction. Since the transitions take place in the condensed RM phase, the problem is quite difficult to solve theoretically. The few processes that are interatomic processes may be possible to calculate by quasiclassical methods, but most processes involve electron transfer possibly over long distances in a metallic or even superconducting material. Such systems cannot be handled by any theoretical methods at present. Thus, no theoretical relation between the intensities of the DIBs and the densities of RM in the interstellar clouds can be found at present. The relative intensities of the various DIBs can neither be determined, since the processes are very different and many details are still unknown. It is probably necessary to form the He(RM) material in the laboratory and attempt to study the electron transfer processes. Unfortunately, He(RM) has not yet been reported, and studies of such a material will probably be quite demanding. Since it is possible to form H(RM) in large quantities and under controlled conditions in the laboratory (Badiei & Holmlid 2006; Holmlid 2008d), it might be possible to measure the DIB absorptions by sensitive laser methods in mixed H-He phases.

## 4.12. Families of DIBs

One of the main ideas to investigate the DIBs has been that they can be divided into families of lines which correlate in intensity and thus may belong to the same carrier (e.g. the same type of molecule) (Krelowski & Walker 1987; Krelowski 1999). It was however recently concluded (Galazutdinov et al. 2003) that families of lines cannot be found reliably. This is in agreement with the results presented here. A detailed comparison will however be made here, partially due to the agreement of one of the families suggested with the emission bands from the Red Rectangle (Scarrott et al. 1992; Miles, Sarre & Scarrott 1995).

Krelowski and Walker (1995) proposed three families of DIBs. The assignments given for the DIBs in these three families are collected in Table 9. One of them consists of only two broad bands 1) 443.0 and 618.0 nm. The other two families are 2) 578.0, 619.6, 620.3, 626.9, and 628.4 nm, and 3) (220), 579.7, 585.0, 637.6, 637.9, and 661.4 nm. The two broad bands in the first family can be found in Table 4 as 442.9 and 617.7 nm. The only property relating these two DIBs given by Krelowski and Walker is that they are missing in the star they study, so not much can be said about the relation of these bands. The second family is more interesting. The assignments in Table 9 show that they all start at doubly excited states 2/1 and 2/2, but the one at 578.0 nm which has the more complex origin in both RM and a singly excited He atom discussed above in subsection 4.4. However, the relation between the RM and the doubly excited atoms is close, and it thus seems that this second family of DIBs has a similar origin in doubly excited states of various atoms, not only He atoms.

The third DIBs family proposed by Krelowski and Walker may have a more varied origin. The band at 220 nm is not discussed at all by us. The assignments in Table 9 show that 579.7 and 585.0 nm are from singly excited He atoms, while the other three DIBs start in doubly excited states. One of these, 661.4 nm, may be a transition within one atom, since the inner electron remains in the same level during the transition. The direct relation between the DIBs in this family appears relatively weak. This is the family which is also observed in emission in the Red Triangle (Scarrott et al. 1992; Miles, Sarre & Scarrott 1995). From the emission point of view, the relation between these DIBs is much stronger, since they all start in doubly excited atomic levels 3/x. Thus, it is possible that a strong correlation will exist in emission between the members of this third DIBs family. The further bands reported in emission by Miles, Sarre & Scarrott (1995) (576.6, 581.8, 582.8 and 591.0 nm) are assigned to be general weak DIBs in Table 10. In emission, they will start in a high Rydberg state or in high doubly excited states and they should thus not be considered to belong to the third DIBs family proposed by Krelowski and Walker.

## 4.13. Recent observations

In this subsection, the DIBs are given in Å to simplify the comparison with the cited references since no recalculated values are given in tables for  this subsection.

Some recent studies have been concerned with DIBs or bands sometimes assigned as DIBs. Munari et al. observed the band 8620 Å and its variation with reddening, and concluded it to be a correct DIB. This band is interpreted in Tables 3 and 4 as an intense transition between two doubly excited states, probably even an intra-atomic transition since the state of the inner electron is unchanged. The agreement with the theoretically calculated value (not adjusted for this purpose) is good. The three other possible DIBs searched for by them, at 8531, 8572 and 8648 Å do not vary correctly with reddening and are concluded by Munari et al. to not be real DIBs. It is not possible to match any of these three bands accurately with transitions of the types investigated here. The best agreement is with the transition $^{3}4/4 \leftarrow {}^{1}2/3$ giving 11650 $cm^{-1}$ instead of the 11663 $cm^{-1}$ determined from the possible band at 8572 Å. This band probably exists, but it is not identified clearly by Munari et al.

The survey study by Hobbs et al. (2009) gives many new DIBs which will be assigned in fortcoming publications from us. One important point in this study is that no preferred wavenumber spacings were found between the DIBs. In this study, not only spacings between adjacent DIBs were tested but also spacings between more than 11 000 pairs of DIBs up to a pair spacing of 400 $cm^{-1}$. No statistically significant pair spacings were detected among the sharp and weak DIBs used. This indicates strongly that the DIBs are not due to any molecules at all. The same conclusion was reached in Hobbs et al. (2008). This means that the probability that the DIBs are due to organic molecules like PAH is very small, since molecular spectra in the visible will contain constant vibrational (and possibly rotational) spacings. As stated in the introduction and in Holmlid (2008a) the DIBs cannot be due to molecules, but the spectra are typical atomic spectra, from atoms in a condensed phase. The calculations here and in Holmlid (2004a, 2008a) prove this conclusively. Hobbs et al. (2009) further find no correlation with the density of $C_2$ molecules in agreement with this.

Misawa et al. (2009) observed three new DIBs, at 9017, 9210 and 9258 Å (11087, 10855, and 10799 $cm^{-1}$). They are asymmetric and may show effects of other absorbers (Misawa et al. 2009). The bands at 10855 and 10799 $cm^{-1}$ are probably due to transitions to the Rydberg limit from $^{1}3/3$ at 10881 - 52 = 10829 $cm^{-1}$ and its sub-bands due to J splitting. See Table 5 for the other transitions of this Rydberg type. The band at 11087 $cm^{-1}$ does not seem to have a similar explanation. It may belong to the group of bands containing DIBs 8648 and 9577 Å, so far with no final assignment within the present model.

The first observations of DIBs in the galaxy M33 were recently made by Cordiner et al. (2008). They report that the DIBs are up to a factor of two stronger per unit reddening than the comparison standard they employed in our galaxy. Since the reddening is due to particle scattering, this would imply that the particulate content in the line-of-sight is small relative to the amount of DIB carriers. Since molecules are not believed to be common in the intergalactic space, this finding would contradict the belief that the DIBs are caused by large organic molecules, a view which is not supported by any facts at all, as noted already in the introduction. The model used here with He and other atoms within RM clouds is compatible with intergalactic space, where such material is believed to exist at considerable densities. For example, the Farady rotation at radio frequencies observed in intergalactic space was explained quite conveniently by a low density of RM in intergalactic space (Badiei and Holmlid 2002c). The required magnetic field is created by the strong magnetic dipoles of the clusters and their interaction to form stacks and filaments, as shown in Fig. 1, which gives enhanced field strength. Thus, the stronger DIBs observed in M33 may be due to absorption in RM in the cloud surrounding the galaxy.

A study of a number of DIBs in the Magellanic clouds (Welty et al. 2006) also concerns objects outside our Galaxy. In this case, the strong DIBs 5780, 5797 and 6284 Å were found to be much weaker than in the Galaxy, relative to reddening and to the densities of Na and K. This is reasonable since the metallicity is lower, which will give lower densities of RM. That the dust-to-gas ratio is lower is in agreement with this. The authors study also a few other DIBs called the $C_2$ DIBs, including DIBs 4964, 4985, 5176, 5419, and 5513 Å. They are found to have higher (almost Galactic) intensities. However, these DIBs are weak and their identities are not very clear, for example 5176 Å is not included either in the DIBs catalog or in Galazutdinov et al. (2000). From the results in the present study it is clear that many DIBs are formed by coincidences, and it is not unlikely that the $C_2$ DIBs cited are anomalous and not of the same RM origin as most DIBs.

The recent high-resolution studies of DIBs by Galazutdinov et al. (2008) give results in good agreement with the calculations presented here. For example, the observation that the strong DIB 5780 Å is unchanged between stars but the weak adjacent 5760, 5763, 5766, 5769, and 5793 do not exist in some cases is confirmed by the assignments done here. Since this spectral region is in the RM range as shown in Table 6, strong variations of the band intensities are expected due to interaction with the RM transitions which will vary in strength quite unrelated to other features due to the He states. As shown in Table 6, several of the weak bands and 5780 Å are amplified by transitions to high RM levels. The complex background of DIB 5780 was discussed above. It is due to the transition $^{0}3/2 \leftarrow 2\ ^{3}P_{2}$ but with added intensity due to the RM levels. The weak bands agree also with transitions of the type 23/1 $\leftarrow$ /5 /1 which give them further intensity, so this spectral region is quite chaotic. The intensity variations are seen in Fig. 4 in Galazutdinov (2008) where the peaks around 5780 Å change in a complex fashion between the two stars studied there. A normalization at 5795 Å instead of 5780 Å shows that 5780, 5776, 5773 and 5795 vary, amplified by the RM transitions, while other features are weak and almost unchanged. Galazutdinov et al. (2008) note that 5795 Å is of the same type as 5780 Å which is in agreement with the (mixed) assignments in Table 6. A full description of the intensity variations observed requires a detailed theoretical study not possible yet, but it is clear that this spectral range is the most complex identified for the DIBs, caused by the overlap with the RM transitions.

In Galazutdinov et al. (2008), one part of a spectrum with no intensity variation between the stars observed is shown in their Fig. 10. According to the assignment given here, the five bands visible there all belong to the same type of transition, namely $3/2 \leftarrow 2\ ^{1}S$. The two strongest at the extreme left and right in the plot are the ones to the singlet and triplet upper states respectively, while two of the other bands (4975 and 4980 Å) correspond to intermediate J split states included (to match other bands) in Table 1 as $^{0}3/2$. The third weak band at 4969 Å probably has a similar origin due to another not previously identified sub-state of 3/2. Thus, these results are in agreement with the assignments given here.

The description of red- and blueshifting of the bands by Galazutdinov et al. (2008) is also quite intriguing. Fig. 12 in this publication will be discussed briefly. From the assignments used here, 5780 and 6284 Å DIBs are due to absorption in He atoms, singly or low doubly excited, even if DIB 5780 is also influenced by RM transitions as described above. They are both redshifted in the spectra in one of the stars studied. However, the weak band DIB 6287 is unchanged. This band has a quite different origin, namely the two-atom transition 8/6 ← /5 /6. It is not limited to He atoms, as the other DIBs in this comparison are due to the singly excited or low doubly excited states involved. Instead, they may involve almost any metal atoms due to the large quantum number states. Thus, the processes may take place in very different environments. With the assignments now given, further such instances may be detected.

## 5. Conclusions

It is shown that practically all DIBs can be assigned to optical transitions involving two-electron co-planar doubly excited atomic states as initial or final state, either He or other two-electron atoms like Ca and other metals especially Na and Li. A fraction of 70% of the total DIBs intensity is due to absorption in doubly excited states, while 30% is caused by absorption in singly excited He atoms. The absorbing doubly excited atoms are in inverted states, while the singly excited He atoms are more thermal. The co-planar states are strongly promoted by the surrounding planar Rydberg Matter (RM). This Rydberg state-promoting environment also gives many transitions to the Rydberg limit and even a long series of transitions to high electronic states in RM. Very few bands can be due to processes in H(RM), and the main influence of the assumed H(RM) phase surrounding He(RM) is to deplete Rydberg ionic states with even quantum numbers.

## Appendix

The main table (Table 10) with the results from the calculations is available on the web as added material. The list of DIBs used there is based on the results by Galazutdinov et al. (2000) (GRMW). Column A in the table contains remarks due to GRMW, while column F contains average signal relative to unity given by these authors. The errors (column D), equivalent widths (column E) and the FWHM (column G) (all in Å) are from Jenniskens and Désert (1994) to simplify the analysis of the results. Vacuum wavenumbers in $cm^{-1}$ are given in column H, with the corresponding FWHM in $cm^{-1}$ in column I. Column J gives the theoretically calculated wavenumbers for the transitions. The notation of the states used in the main table is the same as used in all tables in this study, with the upper state in columns L and M, and the lower state in columns N and O. The multiplicities of the states are given in columns L and N respectively. If the initial state involves two atoms like in some cases in Table 4, columns N and P are used. The specific atom (if conclusive) giving the transition is also included in column P. The publication (P = Holmlid 2004a; M = Holmlid 2008a) where the band was first assigned as given here is included in column Q. An interpretation of the columns is also included at the bottom of Table 10.

More than one assignment of a DIB is often possible, but in general only the one believed to be the most intense contribution is given in Table 10. If a line appears empty to the left, the data in the line above are applicable. This is the case when two assignments are possible. DIBs from GRMW which are not assigned are still included in Table 10 for completeness.

## Acknowledgment

I want to thank Frederic V. Hessman for his interest in the DIBs, encouraging me to finish this study.

Table 1. Energies in $cm^{-1}$ of the second, outermost electron with principal quantum number $n_4$ in a coplanar Rydberg state in RM with the value of $n_3$ given. The quantum defect δ is given in parantheses. The Rydberg constant for He is used. Values in parantheses are not based on any assigned transitions.

| $n_4$ / $n_3$ | /1 | /2 | /3 | /4 | /5 |
|---|---|---|---|---|---|
| $^3$2 | -28075 (0.0313) | -25188 (0.0367) | -18194 (0.01) | -12497 (0.003) | -8812 (0.025) |
| $^1$2 | -27235 (0.0005) | -24720 (0.013) | -17880 (-0.03) | -12363 (-0.038) | -8776 (0) |
| $^3$3 | -12250 (0.0095) | -11978 (0.011) | -10891 (0.03) | -9051 (0.0) | -7160 (0.02) |
| $^0$3 | | -11958 (0.008) -11930 (0.0045) | | | |
| $^1$3 | -12124 (-0.006) | -11893 (0) | -10881 (0) | -9039 (-0.003) | -7149 (0.014) |
| $^3$4 | | -6970 (0.05) | -6739 (0.046) | | |
| $^1$4 | -6855 (0) | (-6805) (0) | (-6599) (0) | -6081 (-0.014) | |
| $^3$5 | | | | -4195 (0.012) | |
| $^1$5 | -4389 (0) | (-4376) (0) | -4320 (0) | -4178 (0) | |
| $^3$6, $^1$6 | | | | -2980 (0) | |

Table 2. Doubly excited levels in wave numbers below the ionization limit calculated from Eq. (1) adjusted for the Rydberg constant for He. *n*” indicates the quantum number for the inner electron.

| *n*”<br>*n*’ | /1 | /2 | /3 | /4 | /5 | /6 | /7 | /8 | /9 | /10 | /11 | /12 | /13 | /14 |
|---|---|---|---|---|---|---|---|---|---|---|---|---|---|---|
| 2 | -27 218.2 | -24 463.2 | -18 128.7 | -12 486.2 | | | | | | | | | | |
| 3 | -12 172.6 | -11 897.2 | -10 872.6 | -9 049.6 | -7 123.1 | -5 549.4 | | | | | | | | |
| 4 | -6 854.3 | -6 803.6 | -6 597.6 | -6 115.8 | -5 369.3 | -4 532.2 | -3 763.4 | -3 121.6 | | | | | | |
| 5 | -4 388.0 | -4 374.3 | -4 318.5 | -4 176.6 | -3 914.1 | -3 540.8 | -3 113.9 | -2 695.2 | -2 319.6 | -1 997.8 | | | | |
| 6 | -3 047.5 | -3 042.7 | -3 023.8 | -2 974.3 | -2 876.1 | -2 718.1 | -2 506.0 | -2 262.4 | -2 014.3 | -1 780.8 | -1 571.0 | -1 387.4 | | |
| 7 | -2 239.1 | -2 237.1 | -2 229.5 | -2 209.5 | -2 168.8 | -2 099.3 | -1 997.0 | -1 865.3 | -1 714.4 | -1 557.3 | -1 404.5 | -1 262.4 | -1 133.8 | |
| 8 | -1 714.4 | -1 713.3 | -1 710.0 | -1 700.9 | -1 682.2 | -1 649.4 | -1 598.8 | -1 529.0 | -1 441.7 | -1 342.3 | -1 237.4 | -1 133.0 | -1 033.4 | -940.9 |
| 9 | -1 354.6 | -1 354.0 | -1 352.3 | -1 347.8 | -1 338.5 | -1 321.9 | -1 295.6 | -1 257.9 | -1 208.1 | -1 147.4 | -1 078.6 | -1 005.5 | -931.6 | -859.6 |
| 10 | -1 097.2 | -1 096.9 | -1 096.0 | -1 093.6 | -1 088.6 | -1 079.6 | -1 065.3 | -1 044.1 | -1 015.3 | -978.5 | -934.6 | -885.2 | -832.4 | -778.5 |
| 11 | -906.8 | -906.8 | -906.3 | -904.9 | -902.1 | -897.0 | -888.8 | -876.5 | -859.3 | -836.8 | -808.9 | -776.1 | -739.4 | -700.2 |
| 12 | -762.0 | -762.0 | -761.7 | -760.9 | -759.2 | -756.2 | -751.2 | -743.8 | -733.3 | -719.2 | -701.4 | -679.7 | -654.6 | -626.6 |
| 13 | -649.2 | -649.2 | -649.0 | -648.5 | -647.5 | -645.6 | -642.5 | -637.8 | -631.2 | -622.2 | -610.6 | -596.2 | -579.1 | -559.4 |
| 14 | -559.8 | -559.9 | -559.7 | -559.4 | -558.7 | -557.5 | -555.5 | -552.5 | -548.2 | -542.3 | -534.6 | -524.9 | -513.2 | -499.4 |
| 15 | -487.7 | -487.7 | -487.6 | -487.4 | -487.0 | -486.2 | -484.8 | -482.8 | -480.0 | -476.0 | -470.8 | -464.2 | -456.0 | -446.3 |
| 16 | -428.7 | -428.6 | -428.5 | -428.4 | -428.1 | -427.5 | -426.7 | -425.3 | -423.3 | -420.6 | -417.0 | -412.4 | -406.7 | -399.8 |
| 17 | -379.7 | -379.7 | -379.6 | -379.5 | -379.3 | -378.9 | -378.3 | -377.4 | -376.0 | -374.1 | -371.6 | -368.3 | -364.2 | -359.2 |

Table 3. Strong transitions for DIBs, observed from Galazutdinov et al. (2000) and calculated. The He Rydberg constant is used. In the T column, either the equivalent width from Jenniskens and Désert (1994) in Å is given or if this is not known, the transmission from Galazutdinov et al. (2000). The width is from Jenniskens and Désert (1994). (G) exists in the spectra but not in the table in Galazutdinov et al. (2000).

| | Observed $\lambda$ (Å) | T | Observed $\tilde{\nu}$ (cm$^{-1}$) | Width (cm$^{-1}$) | Calculated $\tilde{\nu}$ (cm$^{-1}$) | Difference (cm$^{-1}$) | Upper level | Lower level | Comm. |
|---|---|---|---|---|---|---|---|---|---|
| br | 4501.80 | 0.195 | 22207.1 | 12.3 | 22208 | -1 | $^{3}6/4$ | $^{3}2/2$ | |
| | 4726.27 | 0.036 | 21152.4 | 5.6 | 21152 | 0 | $^{1}3/3$ | 2 $^{1}S$ | |
| br | 5487.67 | 0.121 | 18217.6 | 17.9 | 18218 | -0 | $^{3}4/2$ | $^{3}2/2$ | atom |
| | 5494.10 | <0.95 | 18196.3 | | 18196 | 0 | $^{1}3/4$ | $^{1}2/1$ | |
| br | 5508.35 | 0.132 | 18149.2 | 11.5 | 18141 | 8 | Ry | $^{3}2/3$ | atom |
| br | 5609.73 | 0.035 | 17821.2 | 5.4 | 17827 | -6 | Ry | $^{1}2/3$ | atom |
| br | 5705.20 | 0.096 | 17523.0 | 6.8 | 17525 | -2 | Ry | $n$” = 5 | atom |
| | 5780.37 | <0.95 | 17295.1 | | 17294 | 1 | $^{0}3/2$ | 2 $^{3}P_2$ | Overlap RM |
| | 5796.96 | 0.132 | 17245.6 | 2.9 | 17245 | 1 | $^{3}3/2$ | 2 $^{3}P_2$ | |
| | 5849.80 | 0.048 | 17089.9 | 2.9 | 17100 | -10 | $^{1}3/1$ | 2 $^{3}P_2$ | |
| | 6195.96 | 0.061 | 16135.1 | 1.7 | 16139 | -4 | $^{3}3/4$ | $^{3}2/2$ | |
| br | 6203.08 | 0.107 | 16116.6 | 3.1 | 16117 | -0 | $^{0}3/2$ | $^{3}2/1$ | |
| | 6269.75 | 0.076 | 15945.2 | 2.5 | 15951 | -6 | $^{1}3/1$ | $^{3}2/1$ | atom |
| STR | 6283.85 | 0.618 | 15909.4 | 6.6 | 15908 | 1 | $^{3}2/5$ | $^{1}2/2$ | |
| | 6375.95 | 0.026 | 15679.6 | 1.8 | 15681 | -1 | $^{1}3/4$ | $^{1}2/2$ | |
| | 6379.29 | 0.078 | 15671.4 | 1.9 | 15671 | 0 | $^{3}3/4$ | $^{1}2/2$ | |
| STR | 6613.56 | 0.231 | 15116.3 | 2.5 | 15111 | 5 | $^{1}3/1$ | $^{1}2/1$ | atom |
| | 6660.64 | 0.051 | 15009.4 | 1.9 | 15026 | -17 | 3 $^{3}D_3$ | $^{1}2/1$ | |
| | 6810.49 | 0.018 | 14679.2 | 3.2 | 14679 | 0 | $^{3}2/4$ | 2 $^{1}P$ | |
| | 6993.18 | 0.116 | 14295.7 | 2.0 | 14297 | -1 | $^{3}3/3$ | $^{3}2/2$ | |
| STR | 7224.00 | 0.259 | 13838.9 | 2.1 | 13839 | -0 | $^{3}2/3$ | 2 $^{1}S$ | |
| | 7562.2 | 0.087 | 13220.4 | | 13210 | 10 | $^{3}3/2$ | $^{3}2/2$ | Atom (G) |
| KI | 7699 | <0.95 | 12985.1 | | 12979 | 6 | 3 $^{3}D_3$ | $^{3}2/2$ | (G) |
| br | 7915.36 | 0.019 | 12630.2 | 3.0 | 12644 | -14 | 3/6 | $^{3}2/3$ | |
| | 8026.27 | 0.042 | 12455.7 | 1.2 | 12470 | -14 | $^{3}3/1$ | $^{1}2/2$ | |
| | 8439.38 | <0.95 | 11846.0 | | 11840 | 6 | Ry | $^{1}3/2$ | atom |
| br | 8620.79 | 0.125 | 11596.7 | 2.6 | 11595 | 2 | $^{3}2/3$ | $^{3}4/3$ | atom |

Table 4. Broad and intense DIBs from Jenniskens and Désert (1994) and the DIBs catalog (Jenniskens 2009), and calculated from RM theory. All bands with equivalent width > 0.1 Å or FWHM > 10 cm$^{-1}$ are included. The Rydberg constant for He is used. - means uncertain (as in the DIBs catalog) and o means probable.

| | Observed λ (Å) | Eq. width (Å) | Observed $\tilde{\nu}$ (cm$^{-1}$) | Width (cm$^{-1}$) | Calculated $\tilde{\nu}$ (cm$^{-1}$) | Difference (cm$^{-1}$) | Upper level | Lower level | Comment |
|---|---|---|---|---|---|---|---|---|---|
| - | 3980.0 | | 25118.5 | 126.3 | 25188.0 | -69 | Ry | $^{3}$2/2 | atom |
| - | 4000.0 | | 24992.9 | 125.0 | 25063.0 | -70 | $^{3,1}$4/2 | 2 $^{1}$S | |
| - | 4040.0 | | 24745.5 | 122.5 | 24873.0 | -128 | $^{3,1}$3/5 | 2 $^{1}$S | |
| | 4066.0 | 0.282 | 24587.3 | 90.7 | 24720 | -133 | Ry | $^{1}$2/2 | atom |
| o | 4176.5 | 0.427 | 23936.7 | 131.9 | | | | | |
| | 4428.9 | 2.23 | 22572.6 | 61.2 | 22625.0 | -52 | $^{1}$4/3 | 2$^{3}$P | |
| | 4501.8 | 0.195 | 22207.1 | 12.3 | 22208.0 | -1 | $^{3}$6/4 | $^{3}$2/2 | |
| - | 4595.0 | 0.45 | 21756.7 | 132.6 | 21789.0 | -32 | $^{3}$4/5 | $^{1}$2/1 | |
| - | 4665.5 | 0.165 | 21427.9 | 18.8 | | | | | |
| | 4727.1 | 0.087 | 21148.7 | 17.5 | 21154.0 | -5 | $^{1}$4/4 | $^{1}$2/1 | |
| | 4761.7 | 0.623 | 20995.0 | 111.6 | 21005.0 | -10 | $^{3}$4/4 | $^{1}$2/1 | |
| | 4824.0 | 0.452 | 20723.9 | 139.7 | 20799.0 | -75 | /10 /2 | $^{3}$2/2 | |
| | 4881.8 | 0.611 | 20478.5 | 82.7 | 20525.0 | -46 | 5/4 | $^{1}$2/2 | |
| o | 4969.7 | 0.501 | 20116.3 | 136.4 | 20140.0 | -24 | $^{1}$3/2 | 2 $^{1}$S | |
| - | 5039.1 | 0.284 | 19839.3 | 70.5 | 19909.0 | -70 | $^{1}$3/1 | 2 $^{1}$S | |
| | 5109.7 | 0.274 | 19565.2 | 45.2 | 19536.0 | 29 | $^{3,1}$2/4 | 2 $^{1}$S | |
| | 5362.1 | 0.082 | 18644.2 | 20.7 | 18655.0 | -11 | $^{1}$2/5 | /4 /5 | |
| | 5418.9 | 0.08 | 18448.8 | 27.6 | 18449.0 | -0 | $^{3}$4/3 | $^{3}$2/2 | |
| | 5449.6 | 0.244 | 18344.9 | 43.8 | 18392.0 | -47 | $^{1}$3/4 | /4 /4 | |
| o | 5456.0 | 0.134 | 18323.4 | 90.7 | 18382.0 | -59 | $^{3}$3/4 | /4 /4 | |
| | 5487.5 | 0.121 | 18217.6 | 17.9 | 18218.0 | -0 | $^{3}$4/2 | $^{3}$2/2 | atom |
| | 5508.4 | 0.132 | 18149.2 | 11.5 | 18141.0 | 8 | Ry | $^{3}$2/3 | atom |
| | 5535.7 | 0.073 | 18059.5 | 42.4 | 18039.0 | 21 | $^{1}$3/5 | $^{3}$2/2 | |
| | 5537.0 | 0.285 | 18055.3 | 75.0 | 18028.0 | 27 | $^{3}$3/5 | $^{3}$2/2 | |
| | 5704.8 | 0.081 | 17524.2 | 20.3 | 17560.0 | -36 | $^{3}$3/5 | $^{1}$2/2 | |
| | 5779.5 | 0.647 | 17297.7 | 47.9 | 17331 | -33 | $^{1}$3/2 | 2 $^{3}$P | |
| | 5780.6 | 0.579 | 17294.4 | 6.3 | 17295.0 | -1 | $^{0}$3/2 | 2 $^{3}$P | Overlap RM |
| | 5795.2 | 0.117 | 17250.9 | 12.2 | 17246.0 | 5 | $^{3}$3/2 | 2 $^{3}$P | |
| | 5797.1 | 0.132 | 17245.2 | 2.9 | 17 245.0 | 0 | $^{3}$3/2 | 2 $^{3}$P | |
| | 6010.6 | 0.141 | 16632.7 | 9.7 | 16649.0 | -16 | 11/1 | /5 /1 head | |
| | 6045.3 | 0.189 | 16537.2 | 38.3 | 16550 | -13 | $^{1}$3/3 | /4 /3 | |
| | 6177.3 | 0.773 | 16183.8 | 60.3 | 16231.0 | -47 | /7 /2 | $^{3}$2/2 | |
| | 6203.2 | 0.107 | 16116.3 | 3.1 | 16117.0 | -1 | $^{0}$3/2 | $^{3}$2/1 | |
| | 6204.3 | 0.189 | 16113.4 | 10.1 | 16117.0 | -4 | $^{0}$3/2 | $^{3}$2/1 | |
| | 6207.8 | 0.136 | 16104.3 | 30.4 | 16097.0 | 7 | $^{3}$3/2 | $^{3}$2/1 | |
| | 6281.1 | 1.237 | 15916.4 | 21.5 | 15908.0 | 8 | $^{3}$2/5 | $^{1}$2/2 | |
| | 6284.3 | 0.618 | 15908.3 | 6.6 | 15908.0 | 0 | $^{3}$2/5 | $^{1}$2/2 | |
| | 6315.0 | 0.352 | 15830.9 | 53.2 | 15884.0 | -53 | /6 /1 | $^{3}$2/1 | |
| | 6359.5 | 0.536 | 15720.2 | 91.5 | 15 763.0 | -43 | /7 /2 | $^{1}$2/2 | |
| | 6413.5 | 0.085 | 15587.8 | 19.7 | | | | | |
| | 6451.6 | 0.403 | 15495.7 | 60.1 | 15538.0 | -42 | $^{1}$3/2 | /4 /2 | |
| | 6494.2 | 0.2 | 15394.1 | 26.1 | 15453.0 | -59 | $^{3}$3/2 | /4 /2 | |
| | 6532.1 | 0.664 | 15304.8 | 39.8 | 15316.0 | -11 | 7/1 | /5 /1 head | |
| | 6591.4 | 0.087 | 15167.1 | 12.9 | 15181.0 | -14 | $^{3}$3/1 | /4 / 2 | |
| | 6613.7 | 0.231 | 15116.0 | 2.5 | 15111.0 | 5 | $^{1}$3/1 | $^{1}$2/1 | atom |
| | 6939.0 | 0.396 | 14407.3 | 43.6 | 14489.0 | -82 | 3 $^{3}$P | $^{1}$2/1 | |
| | 6993.2 | 0.116 | 14295.7 | 2.0 | 14297.0 | -1 | $^{3}$3/3 | $^{3}$2/2 | |
| o | 7223.1 | 0.083 | 13840.7 | 10.4 | | | | | |
| | 7224.2 | 0.259 | 13838.5 | 2.1 | 13839.0 | -0 | $^{1}$3/3 | $^{1}$2/2 | |
| | 7357.2 | 0.227 | 13588.4 | 51.7 | 13560.0 | 28 | $^{1}$5/3 | $^{1}$2/3 | |
| o | 7398.6 | 0.12 | 13512.3 | 2.0 | | | | | |
| | 7432.1 | 0.549 | 13451.4 | 39.8 | 13491.0 | -40 | /10 /3 | $^{1}$2/3 | |
| o | 7569.7 | 0.216 | 13206.9 | 9.6 | 13167.0 | 40 | $^{3}$5/1 | /5 /1 head | |
| | 7709.7 | 0.444 | 12967.1 | 57.2 | 12997 | -30 | /6 /2 | $^{3}$2/2 | |
| | 7927.8 | 0.428 | 12610.4 | 23.9 | 12596 | 14 | $^{1}$3/1 | $^{1}$2/2 | |
| | 8621.1 | 0.125 | 11596.3 | 2.6 | 11595.0 | 1 | $^{1}$4/3 | $^{3}$2/3 | atom |
| | 8621.2 | 0.272 | 11596.1 | 7.5 | 11595.0 | 1 | $^{1}$4/3 | $^{3}$2/3 | atom |
| | 8648.3 | 0.241 | 11559.8 | 5.6 | | | | | |
| | 9577.0 | 0.398 | 10438.8 | 4.5 | | | | | |
| o | 9632.0 | 0.573 | 10379.2 | 4.3 | 10380.0 | -1 | $^{3}$2/1 | 2 $^{3}$S | |
| | 11797.0 | 0.13 | 8474.4 | 1.9 | 8475.0 | -1 | $^{1}$4/3 | 3 $^{3}$S | |
| | 13175.0 | 0.36 | 7588.1 | 2.3 | 7589.0 | -1 | /10 /2 | $^{3}$3/2 | |

Table 5. Transitions from the low levels indicated, to the Rydberg and RM limits. Only transitions within the ranges of the DIBs surveys are included. JD indicates Jenniskens and Désert (1994), the other DIBs are from Galazutdinov et al. (2000).

| Level | Calculated $\tilde{\nu}$ ($cm^{-1}$) | Observed $\tilde{\nu}$ ($cm^{-1}$) | Difference ($cm^{-1}$) | FWHM ($cm^{-1}$) | λ (Å) | Comment |
|---|---|---|---|---|---|---|
| $^{3}2/2$ | 25188 | 25118.5 | 70 | 126.3 | 3980 | JD |
| $^{3}2/3$ | 18194 | 18149.2 | 45 | 11.5 | 5508.35 | |
| $^{3}2/4$ | 12497 | 12437.6 | 59 | 1.7 | 8037.9 | |
| $^{1}2/2$ | 24720 | 24587.3 | 133 | 90.7 | 4066.0 | JD |
| $^{1}2/3$ | 17880 | 17821.2 | 59 | 5.4 | 5609.7 | |
| $^{1}2/4$ | 12363 | 12303.2 | 60 | - | 8125.75 | |
| $^{3}3/1$ | 12250 | | | | 8161.0 | Possible JD |
| $^{3}3/2$ | 11978 | | | | 8346.5 | Possible JD |
| $^{0}3/2$ | 11958<br>11930 | | | | 8360.0<br>8380.0 | Possible JD |
| $^{1}3/1$ | 12124 | 12069.2 | 55 | 1.7 | 8283.29 | |
| $^{1}3/2$ | 11893 | 11846.0 | 47 | - | 8439.38 | |
| /5 | 17555 | 17523.0 | 32 | 6.8 | 5705.2 | |
| | | | | | | |
| $3\ ^{3}S$ | 15074 | 15072.3 | 2 | 2.5 | 6632.85 | |
| $3\ ^{1}S$ | 13446 | 13451.4 | -5 | 39.8 | 7432.1 | JD |
| $3\ ^{3}P$ | 12746 | 12745.3 | 1 | 7.8 | 7843.9 | |

Table 6. Transitions to high RM levels from *n*” = 5. *n*” is the principal quantum number at an ion ($Z = 2$). GMKW indicates Galazutdinov et al. (2000). Overlaps and coincidences are indicated in the Comment column.

| *n*” | Calculated $\tilde{\nu}$ (cm$^{-1}$) | Observed $\tilde{\nu}$ (cm$^{-1}$) | Diffe-rence (cm$^{-1}$) | Observed λ (Å) | Comment |
|---|---|---|---|---|---|
| 48 | 17365.1 | - | | (5757) | |
| 47 | 17356.9 | 17355.1 | 1.8 | 5760.40 | |
| 46 | 17348.1 | 17348.2 | -0.1 | 5762.70 | Coincidence 23/1 ← /5 /1 |
| 45 | 17338.8 | 17337.8 | 1.0 | 5766.16 | |
| 44 | 17328.9 | 17329.1 | -0.2 | 5769.04 | Coincidence 22/1 ← /5 /1 |
| 43 | 17318.2 | 17318.4 | -0.2 | 5772.6 | |
| 42 | 17306.8 | 17308.9 | -2.1 | 5775.78 | Coincidence 21/1 ← /5 /1 |
| 41 | 17294.5 | 17295.1 | -0.6 | 5780.37 | overlap $^{0}3/2 \leftarrow 2\ ^{3}P$ |
| 40 | 17281.2 | 17281.1 | 0.1 | 5785.05 | Coincidence 20/1 ← /5 /1 |
| 39 | 17267.0 | 17269.7 | -2.7 | 5788.90 | Weselak et al. (2000) |
| 38 | 17251.6 | 17251.0 | 0.6 | 5795.16 | overlap |
| 37 | 17235.0 | 17232 | 3 | 5801.5 | Peak GMKW |
| 36 | 17216.9 | 17216.8 | 0.1 | 5806.68 | |
| 35 | 17197.3 | 17201.1 | -3.8 | 5811.96 | |
| 34 | 17175.9 | 17177 | -1 | 5820 | Small peak GMKW |
| 33 | 17152.5 | 17152.4 | 0.1 | 5828.46 | |
| 32 | 17127.0 | 17124.4 | 2.6 | 5838 | |
| 31 | 17098.9 | 17104.5 | -5.6 | 5844.8 | overlap |
| 30 | 17067.9 | - | | (5857) | |

Table 7. Predicted positions of DIBs in wave numbers due to transitions x/y ← 2 $^3$S where the lower singly excited state is at 38455 cm$^{-1}$ below the ionization limit. The levels for the doubly excited states are given in Table 1. The predicted corrected (air) wavelengths in Å are given in parentheses. The only band observed so far is underlined.

| Level | /1 | /2 | /3 | /4 | /5 |
|---|---|---|---|---|---|
| $^3$2 | <u>10380 (9631.3)</u> | 13267 (7535.4) | 20261 (4934.2) | 25958 (3851.3) | 29643 (3372.5) |
| $^1$2 | 11220 (8910.2) | 13735 (7278.7) | 20575 (4859.0) | 26092 (3831.5) | 29679 (3368.4) |
| $^3$3 | 26205 (3815.0) | 26477 (3775.8) | 27564 (3626.9) | 29406 (3399.7) | 31295 (3194.5) |
| $^1$3 | 26331 (3796.7) | 26562 (3763.7) | 27574 (3625.6) | 29416 (3398.5) | 31306 (3193.4) |
| $^3$4 | | 31485 (3175.2) | 31716 (3152.1) | | |
| $^1$4 | 31600 (3163.6) | 31650 (3158.6) | 31856 (3138.2) | 32374 (3088.0) | |

Table 8. Summed intensities for most important initial states in DIBs. “Intense” refers to DIBs in Table 3 and “Broad” to those in Table 4. The values given are summed equivalent widths in Å from Jenniskens and Désert (1994) and Jenniskens (2009).

| Intense | /1 | /2 | /3 | Sum |
|---|---|---|---|---|
| $^{3}2$ | 1.18 | 0.72 | 0.33 | 2.22 |
| $^{1}2$ | 0.42 | 1.17 | 0.44 | 2.02 |
| Broad | | | | |
| $^{3}2$ | 1.37 | 1.77 | 0.53 | 3.67 |
| $^{1}2$ | 1.41 | 3.60 | 0.36 | 5.37 |
| | | | | |
| Intense | 2 S | 2 P | 3 S | |
| $^{3}$ | | 0.32 | 0.10 | 0.42 |
| $^{1}$ | 0.35 | 0.07 | | 0.42 |
| Broad | | | | |
| $^{3}$ | 0.57 | 4.08 | 0.13 | 4.78 |
| $^{1}$ | 1.06 | | 0.55 | 1.61 |

Table 9. A compilation of the assignments for the three DIBs families proposed by Krelowski and Walker (1987). The He Rydberg constant is used. In the T column, either the equivalent width from Jenniskens and Désert (1994) in Å is given or if this is not known, the transmission from Galazutdinov et al. (2000).

| | Observed λ (Å) | T | Observed $\tilde{\nu}$ (cm$^{-1}$) | Width (cm$^{-1}$) | Calculated $\tilde{\nu}$ (cm$^{-1}$) | Difference (cm$^{-1}$) | Upper level | Lower level | Comm. |
|---|---|---|---|---|---|---|---|---|---|
| | Family 1 | | | | | | | | |
| | 4428.9 | 2.23 | 22572.6 | 61.2 | 22625.0 | -52 | $^{1}4/3$ | $2^{3}P$ | |
| | 6177.3 | 0.773 | 16183.8 | 60.3 | 16231.0 | -47 | /7 /2 | $^{3}2/2$ | |
| | | | | | | | | | |
| | Family 2 | | | | | | | | |
| | 5780.37 | <0.95 | 17295.1 | | 17294 | 1 | $^{0}3/2$ | $2\,^{3}P_2$ | Overlap RM |
| | 6195.96 | 0.061 | 16135.1 | 1.7 | 16139 | -4 | $^{3}3/4$ | $^{3}2/2$ | |
| br | 6203.08 | 0.107 | 16116.6 | 3.1 | 16117 | -0 | $^{0}3/2$ | $^{3}2/1$ | |
| | 6269.75 | 0.076 | 15945.2 | 2.5 | 15951 | -6 | $^{1}3/1$ | $^{3}2/1$ | atom |
| STR | 6283.85 | 0.618 | 15909.4 | 6.6 | 15908 | 1 | $^{3}2/5$ | $^{1}2/2$ | |
| | | | | | | | | | |
| | Family 3 | | | | | | | | |
| | 5796.96 | 0.132 | 17245.6 | 2.9 | 17245 | 1 | $^{3}3/2$ | $2\,^{3}P_2$ | |
| | 5849.80 | 0.048 | 17089.9 | 2.9 | 17100 | -10 | $^{1}3/1$ | $2\,^{3}P_2$ | |
| | 6375.95 | 0.026 | 15679.6 | 1.8 | 15681 | -1 | $^{1}3/4$ | $^{1}2/2$ | |
| | 6379.29 | 0.078 | 15671.4 | 1.9 | 15671 | 0 | $^{3}3/4$ | $^{1}2/2$ | |
| STR | 6613.56 | 0.231 | 15116.3 | 2.5 | 15111 | 5 | $^{1}3/1$ | $^{1}2/1$ | atom |

## Figure captions

Fig. 1. A stack of RM clusters $He_7$.

Fig. 2. Angular momentum couplings in RM clusters, in the dark state in stacks to the left and free to the right. $\underline{R}$ is the rotational angular momentum of the cluster, with its components $\underline{K}_b$ and $\underline{K}_c$ where $\underline{K}_c$ is along the figure axis of the planar six-fold symmetric cluster. The electron orbital angular momentum is $\underline{L}$, in the direction opposite to the electron spin quantum number $\underline{S}$, these two giving $\underline{\Omega}$, if they are coupled. Finally the total angular momentum $\underline{J}$ is formed from $\underline{\Omega}$ and $\underline{R}$.

Fig. 3. Electron excitations and transfer processes involved in sharp and intense DIB absorption bands starting from co-planar doubly excited atomic states. In panel A, an example of an intra-atomic transition is shown. In panel B, an electron transfer is shown.

Fig. 4. Electron transfer processes involved in many broad DIBs. In panel A, the Fermi level varies with distance in the RM material. In panel B, the two electrons are coupled through the conduction band. See the text.

Fig. 5. Electron transfer processes involved in other broad and weak DIBs. Panel A shows a process starting in a doubly excited state, while panel B instead ends in a doubly excited state.

Fig. 6. Energy diagram for the co-planar doubly excited states with transitions drawn for the sharp and intense bands assigned in Table 3.

Fig. 7. Energy diagram for the co-planar doubly excited states with transitions drawn for the broad bands assigned in Table 4.

Fig. 8. Energy diagram for the co-planar doubly excited states with transitions drawn for the Rydberg and RM related bands assigned in Table 5.

Fig. 9. Some of the most intense DIBs interpreted here. Wavelengths in bold and italics indicate doubly excited atoms as the initial state, while normal type wavelengths indicate singly excited He atoms as the initial state. The band in a box is of mixed special origin partly due to a transition to an RM level, see text. The synthesized DIB spectrum is from Sorokin et al. (1998).

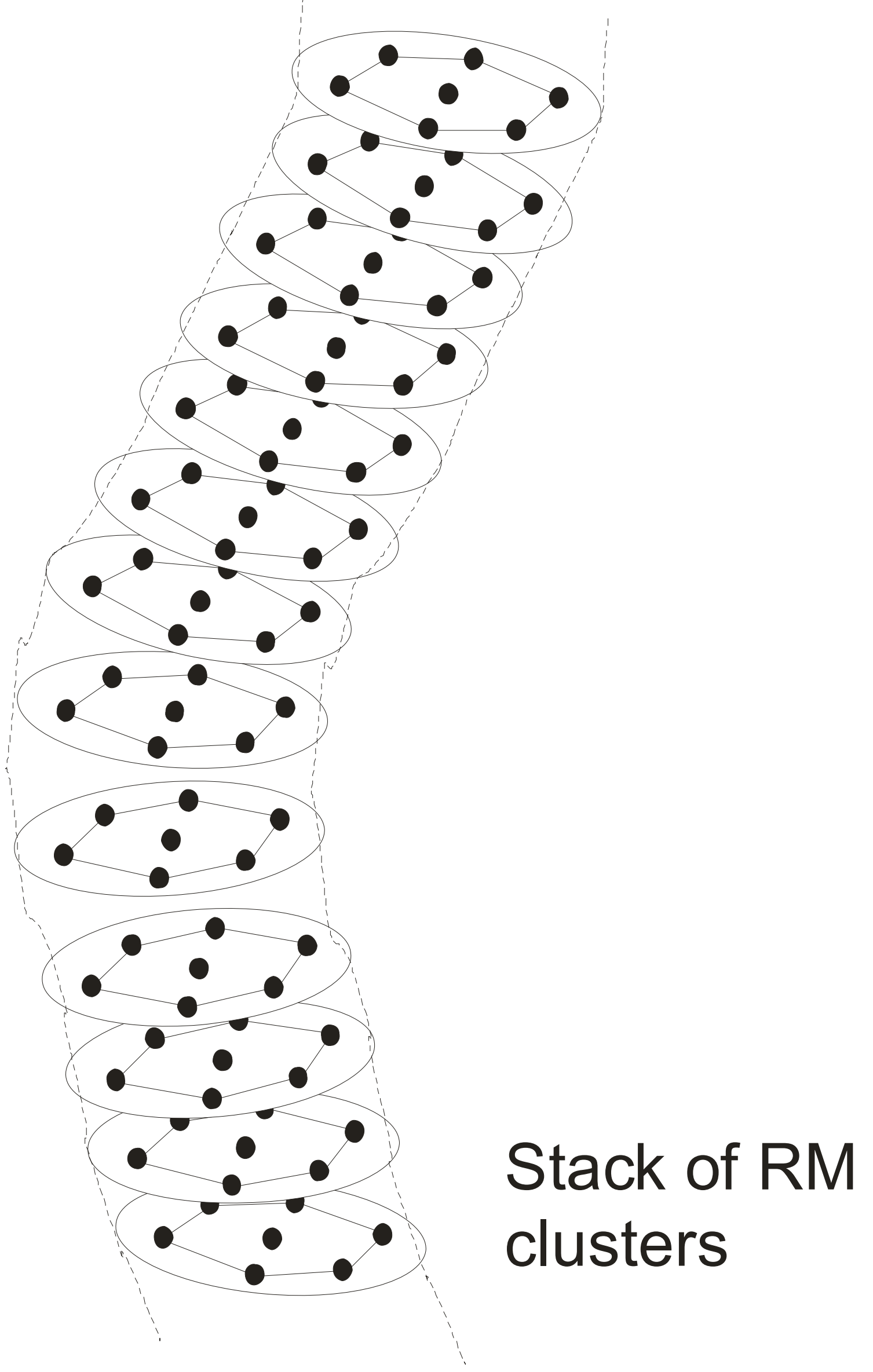

Fig. 1.

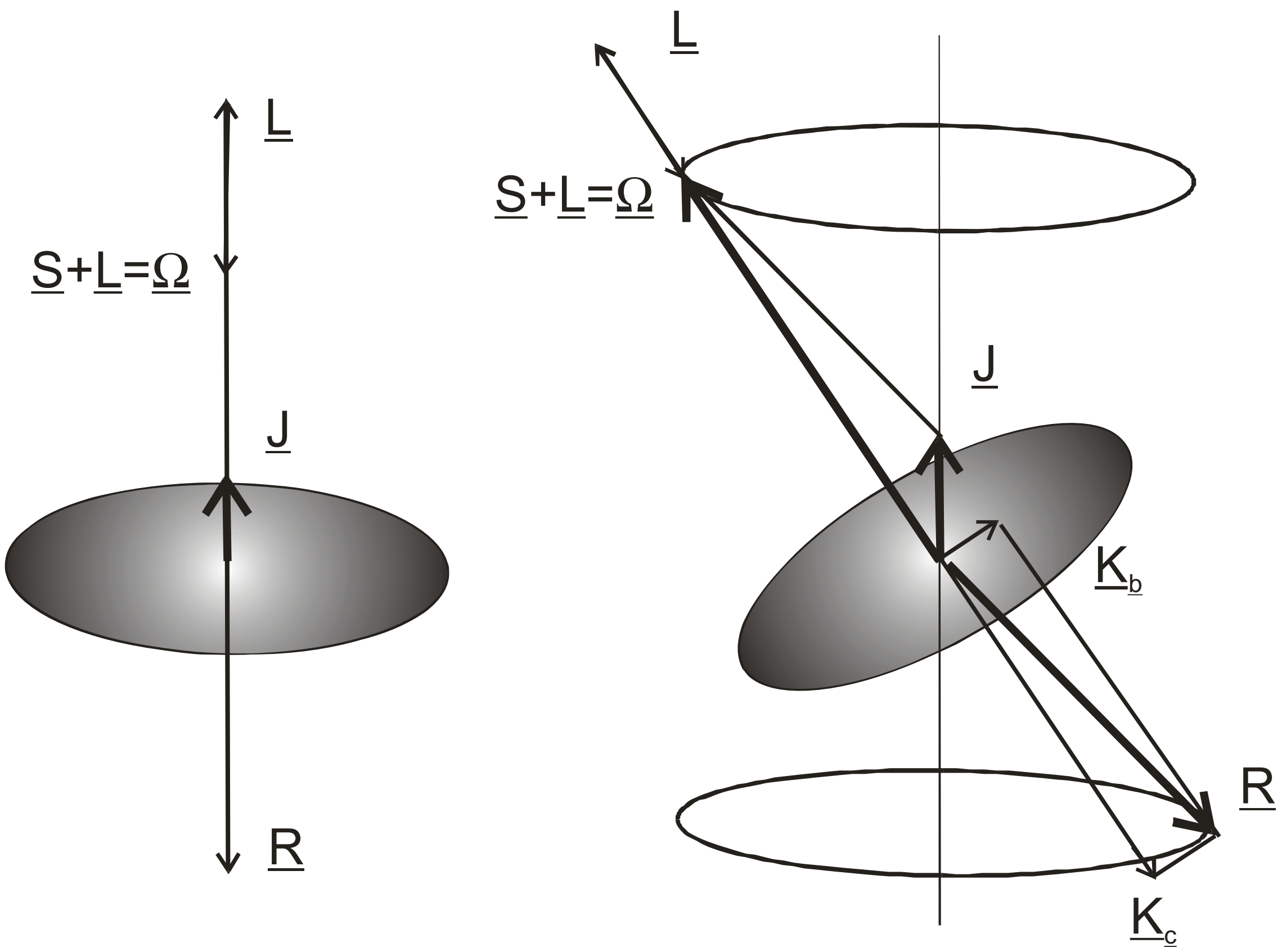

Fig. 2

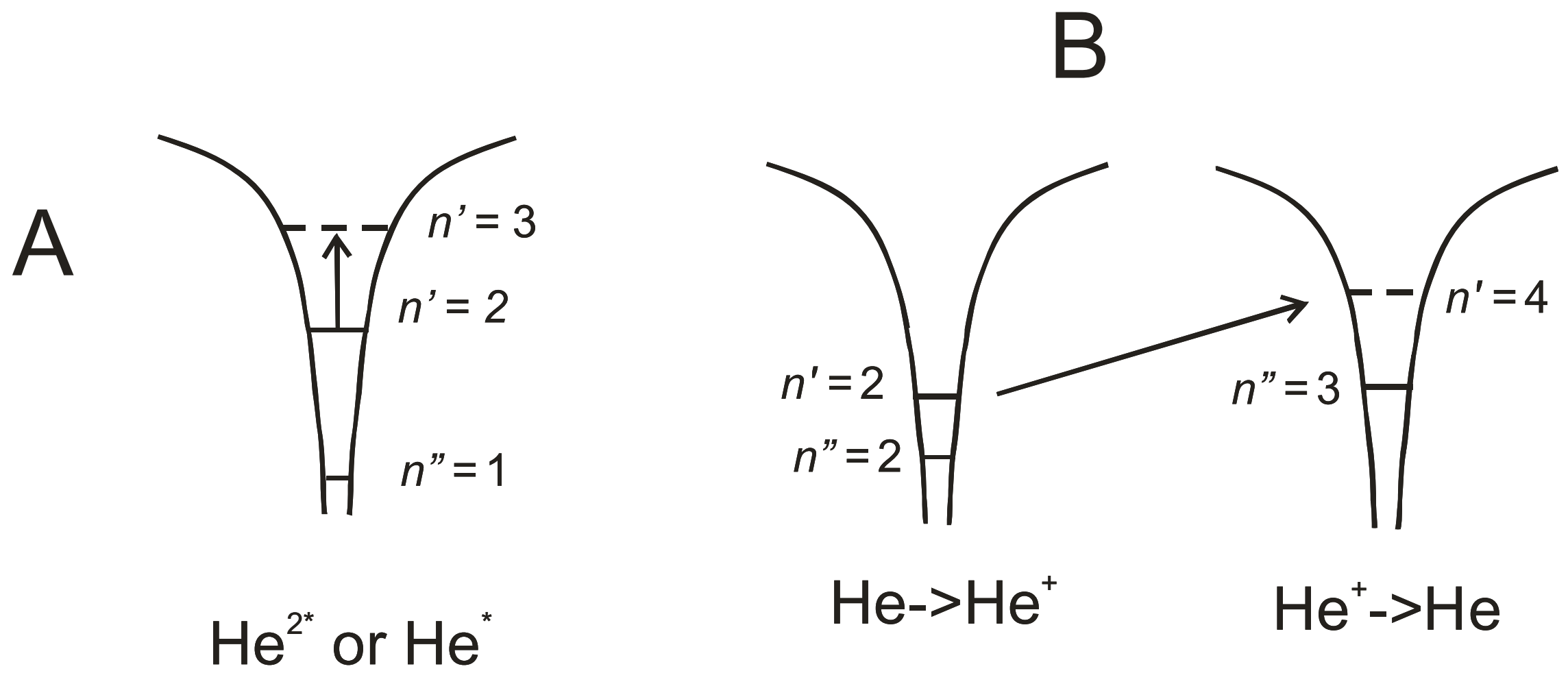

Fig. 3

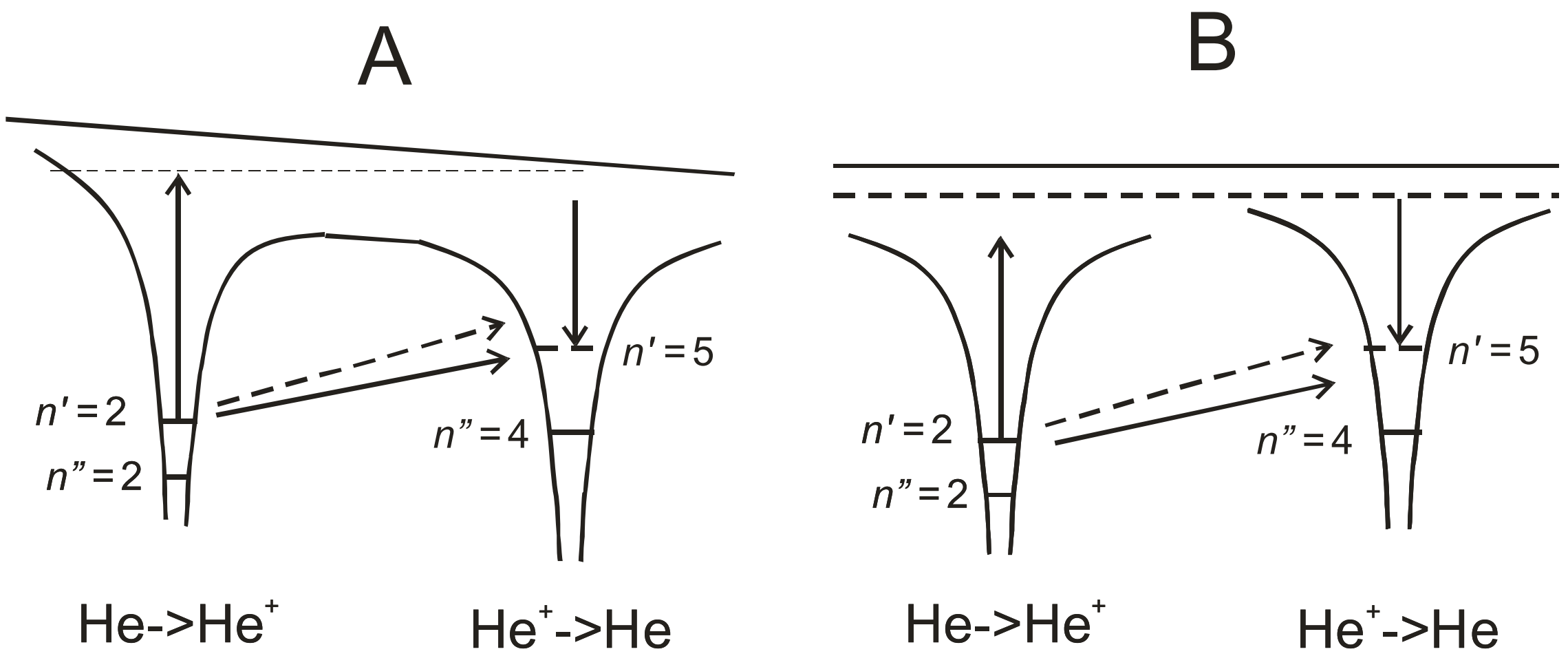

Fig. 4

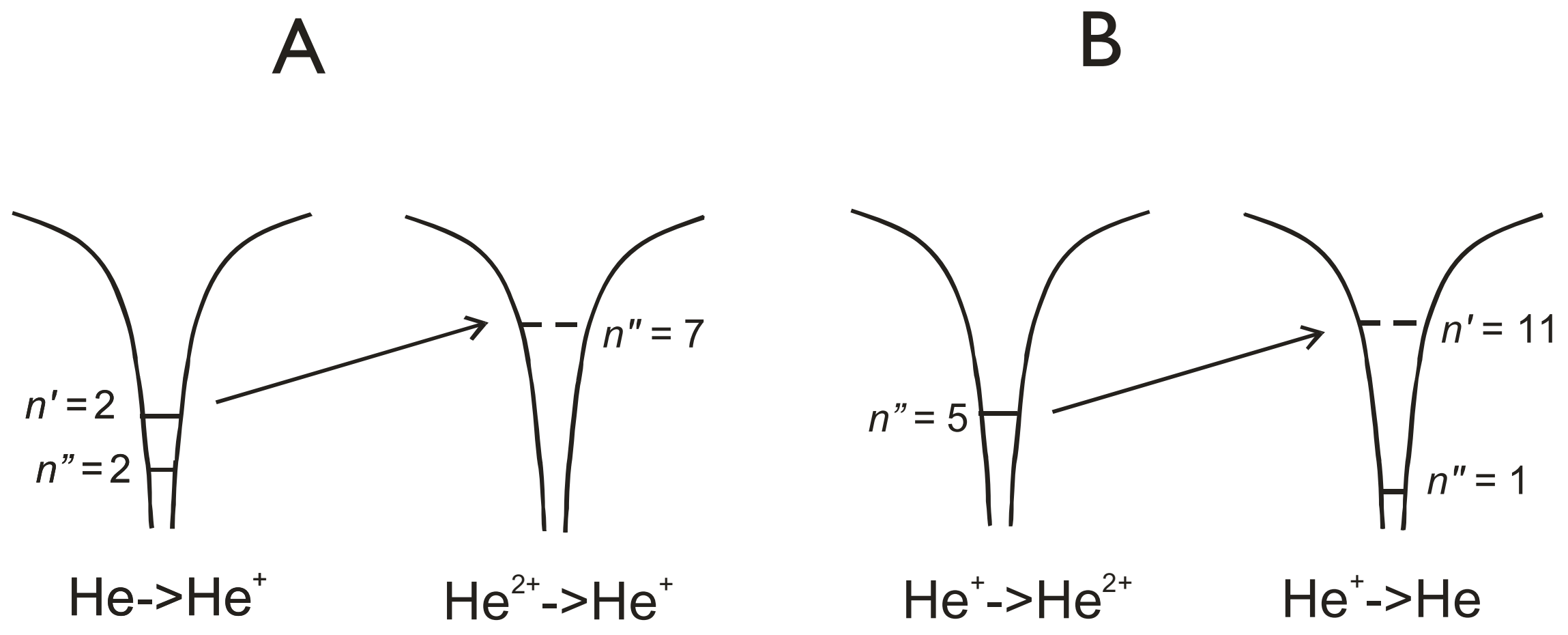

Fig. 5

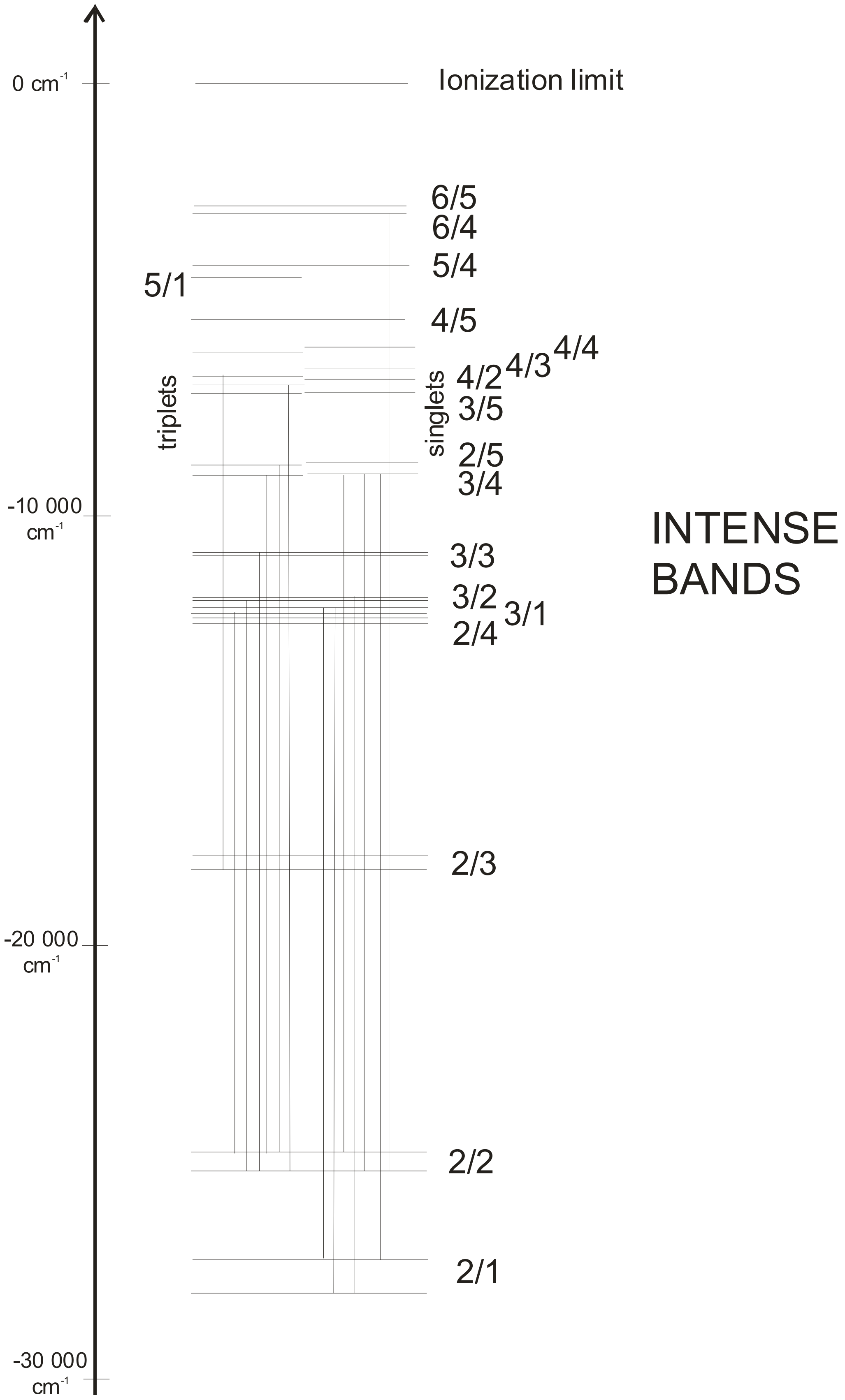

Fig. 6

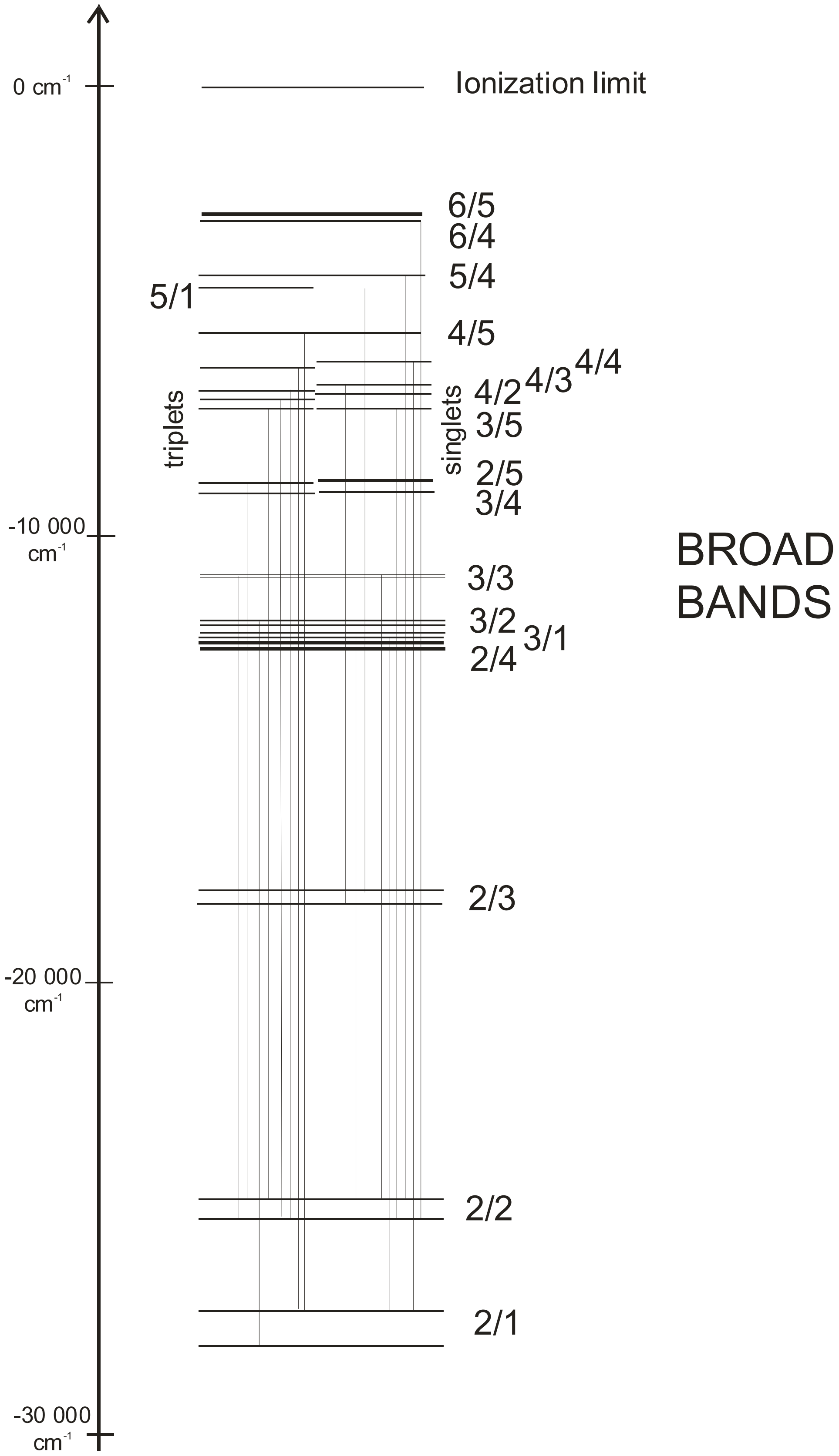

Fig. 7

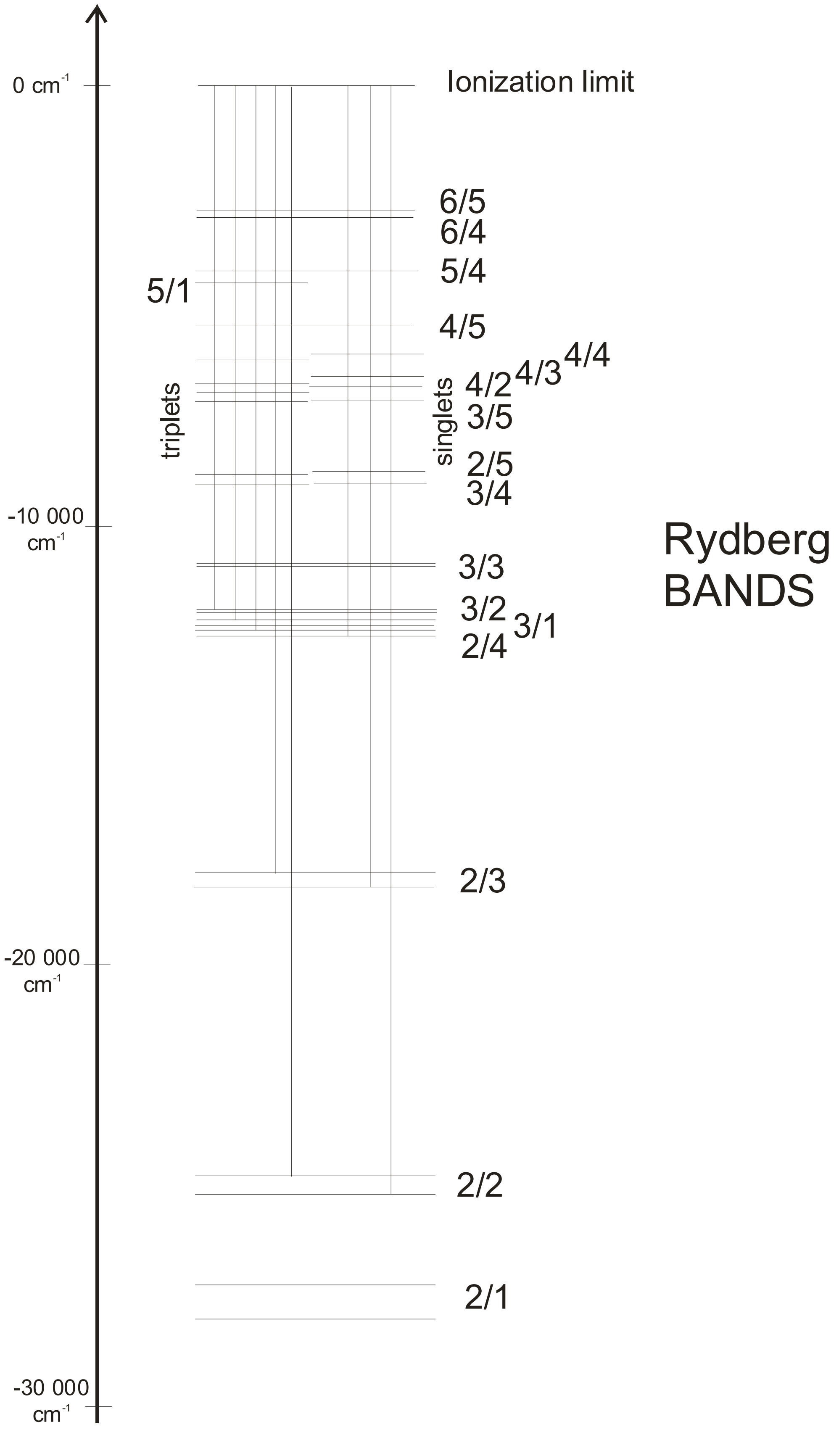

Fig. 8

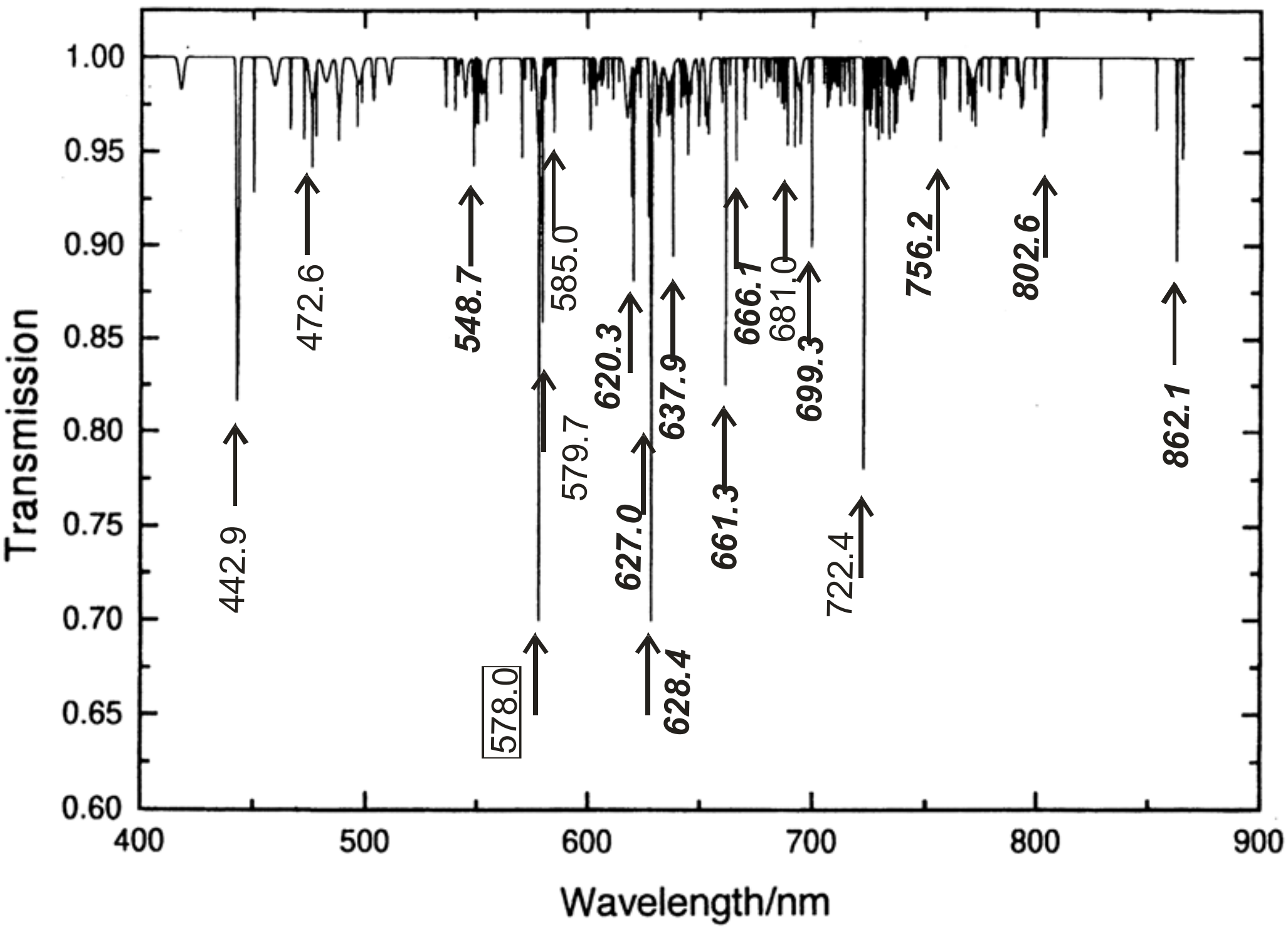

Fig. 9

| **A** | **A** | **B** | **C** | **D** | **E** | **F** | **G** | **H** | **I** | **J** | **K** | **L** | **M** | **N** | **O** | **P** | **Q** | **R** |
|---|---|---|---|---|---|---|---|---|---|---|---|---|---|---|---|---|---|---|
| **1** | | | | | | | | | | | | | | | | | | |
| **2** | com | lambda | | error | width | abs | FWHM | ny | FWHM | ny calc. | delta | | upper | | lower | atom | Ref. | |
| **3** | br | 4 501.80 | | 0.7 | 0.195 | <0.95 | 2.5 | 22207.1 | 12.3 | 22208.0 | -1 | 3 | 6/4 | 3 | 2/2 | | new | |
| **4** | | 4 726.27 | | 0.07 | 0.036 | <0.95 | 1.25 | 21152.4 | 5.6 | 21152.0 | 0 | 1 | 3/3 | | 2 $^1$S | **He** | M | |
| **5** | | 4 762.67 | | 0.32 | 0.079 | | 2.1 | 20990.8 | 9.3 | 20993.0 | -2 | 3 | 5/4 | 3 | 2/2 | | new | |
| **6** | | | | | | | | | | 20981.0 | 10 | | 4 $^3$P | 3 | 2/1 | **He** | new | |
| **7** | | 4 780.04 | | 0.29 | 0.057 | | 1.45 | 20914.5 | 6.3 | 20915.0 | -1 | 3 | 3/5 | 3 | 2/1 | | new | |
| **8** | | 4 963.90 | | 0.04 | 0.016 | | 0.63 | 20139.8 | 2.6 | 20140.0 | -0 | 1 | 3/2 | | 2 $^1$S | **He** | M | |
| **9** | | | | | | | | | | 20141.0 | -2 | | 4 $^3$P | 1 | 2/1 | **He** | new | |
| **10** | | 4 984.81 | | 0.12 | 0.014 | | 0.65 | 20055.3 | 2.6 | 20055.0 | 0 | 3 | 3/2 | | 2 $^1$S | **He** | M | |
| **11** | | 5 363.60 | | 0.14 | 0.032 | | 1.99 | 18639.0 | 6.9 | 18637.0 | 2 | 3,1 | 2/5 | | 5 | 4 | new | |
| **12** | | 5 404.50 | | 0.26 | 0.038 | | 1.3 | 18498.0 | 4.5 | 18490.0 | 8 | 1 | 4/4 | 1 | 2/2 | | new | |
| **13** | | 5 418.90 | | 0.66 | 0.08 | | 8.1 | 18448.8 | 27.6 | 18449.0 | -0 | 3 | 4/3 | 3 | 2/2 | | new | |
| **14** | br | 5 487.67 | | 0.43 | 0.121 | <0.95 | 5.4 | 18217.6 | 17.9 | 18218.0 | -0 | 3 | 4/2 | 3 | 2/2 | | new | |
| **15** | | 5 494.10 | | | | <0.95 | | 18196.3 | | 18196.0 | 0 | 1 | 3/4 | 1 | 2/1 | | M | |
| **16** | br | 5 508.35 | | 0.03 | 0.132 | | 3.5 | 18149.2 | 11.5 | 18141.0 | 8 | Ry | | 3 | 2/3 | | M | |
| **17** | | 5 512.64 | | | | | | 18135.1 | | 18127.0 | 8 | 3 | 3/4 | | 2 $^1$P | **He** | M | |
| **18** | | 5 541.62 | | 1.1 | 0.015 | | 1.6 | 18040.3 | 5.2 | 18039.0 | 1 | 1 | 3/5 | 3 | 2/2 | | new | |
| **19** | | 5 544.96 | | 0.05 | 0.02 | | 0.79 | 18029.4 | 2.6 | 18039.0 | -10 | 1 | 3/5 | 3 | 2/2 | | new | |
| **20** | | 5 546.46 | | | | | | 18024.5 | | 18028.0 | -3 | 3 | 3/5 | 3 | 2/2 | | new | |
| **21** | | 5 594.59 | | | | | | 17869.4 | | 17865.0 | 4 | 1 | 4/1 | 1 | 2/2 | | new | |
| **22** | br | 5 609.73 | | 0.23 | 0.035 | | 1.7 | 17821.2 | 5.4 | 17827.0 | -6 | Ry | | 1 | 2/3 | | M | |
| **23** | br | 5 705.20 | | 0.11 | 0.096 | | 2.2 | 17523.0 | 6.8 | 17525.0 | -2 | Ry | | | /5 | | new | |
| **24** | | 5 719.30 | | 0.1 | 0.012 | | 0.93 | 17479.8 | 2.8 | | | | | | | | | |
| **25** | | 5 760.40 | | 0.2 | 0.002 | | 0.7 | 17355.1 | 2.1 | 17357.0 | -2 | RM | /47 | | /5 | | M | |
| **26** | | 5 762.70 | | 0.33 | 0.008 | | 0.66 | 17348.2 | 2.0 | 17348.0 | 0 | RM | /46 | | /5 | | M | |
| **27** | | 5 766.16 | | 0.14 | 0.013 | | 0.92 | 17337.8 | 2.8 | 17339.0 | -1 | RM | /45 | | /5 | | M | |
| **28** | | 5 769.04 | | 0.2 | 0.003 | | 0.8 | 17329.1 | 2.4 | 17329.0 | 0 | RM | /44 | | /5 | | M | |
| **29** | | | | | | | | | | 17331.0 | -2 | 1 | 3/2 | | 2 $^3P_2$ | **He** | M | |
| **30** | | 5 772.60 | | 0.04 | 0.02 | | 1 | 17318.4 | 3.0 | 17318.0 | 0 | RM | /43 | | /5 | | M | |
| **31** | | 5 775.78 | | 0.23 | 0.01 | | 0.85 | 17308.9 | 2.5 | 17307.0 | 2 | RM | /42 | | /5 | | M | |
| **32** | | 5 780.37 | | 0.1 | 0.579 | <0.95 | 2.07 | 17295.1 | 6.2 | 17295.0 | 0 | RM | /41 | | /5 | | M | |
| **33** | | | | | | | | | | 17294.0 | 1 | 0 | 3/2 | | 2 $^3P_2$ | **He** | new | |
| **34** | | 5 785.05 | | 0.25 | 0.018 | | 1 | 17281.1 | 3.0 | 17281.2 | -0 | RM | /40 | | /5 | | M | |
| **35** | | 5 793.22 | | 0.2 | 0.012 | | 0.94 | 17256.8 | 2.8 | 17258.4 | -2 | | 18/19 | | /5 | /19 | new | |
| **36** | | 5 795.16 | | 0.32 | 0.117 | | 4.1 | 17251.0 | 12.2 | 17252.0 | -1 | RM | /38 | | /5 | | M | |
| **37** | | 5 796.96 | | 0.05 | 0.132 | <0.95 | 0.97 | 17245.6 | 2.9 | 17246.0 | -0 | 3 | 3/2 | | 2 $^3P_2$ | **He** | M | |
| **38** | | 5 806.68 | | | | | | 17216.8 | | 17217.0 | -0 | RM | /36 | | /5 | | new | |
| **39** | | 5 809.24 | | 0.06 | 0.016 | | 1.1 | 17209.2 | 3.3 | 17206.7 | 2 | | 17/16 | | /5 | /16 | P | |
| **40** | | 5 811.96 | | | | | | 17201.1 | | 17199.9 | 1 | | 17/15 | | /5 | /15 | new | |
| **41** | | 5 815.71 | | | | | | 17190.0 | | 17189.0 | 1 | | 17/13 | | /5 | /13 | new | |
| **42** | | | | | | | | | | 17194.0 | -4 | 1 | 3/3 | 3 | 2/1 | | M | |
| **43** | | 5 818.75 | | 0.3 | 0.004 | | 0.49 | 17181.1 | 1.4 | 17181.0 | 0 | | 17/11 | | /5 | /11 | P | |
| **44** | | | | | | | | | | 17184.0 | -3 | 3 | 3/3 | 3 | 2/1 | | M | |
| **45** | | 5 821.23 | | | | | | 17173.7 | | 17173.0 | 1 | | 17/16 | | /5 | /1 | new | |
| **46** | | 5 828.46 | | 0.2 | 0.003 | | 0.31 | 17152.4 | 0.9 | 17156.0 | -4 | | 16/14 | | /5 | /14 | P | |
| **47** | | | | | | | | | | 17153.0 | -1 | RM | /33 | | /5 | | new | |
| **48** | | 5 838.00 | | 0.2 | 0.001 | | 0.52 | 17124.4 | 1.5 | 17126.9 | -2 | | 16/1 | | /5 | /1 | P | |
| **49** | | | | | | | | | | 17127.0 | -3 | RM | /32 | | /5 | | new | |
| **50** | | 5 840.65 | | 0.2 | 0.003 | | 0.5 | 17116.6 | 1.5 | 17121.0 | -4 | | 15/15 | | /5 | /15 | P | |
| **51** | | 5 842.23 | | 0.2 | 0.005 | | 0.24 | 17112.0 | 0.7 | 17109.0 | 3 | | 15/14 | | /5 | /14 | new | |
| **52** | | 5 844.80 | | 0.2 | 0.009 | | 0.47 | 17104.5 | 1.4 | 17100.0 | 4 | | 15/13 | | /5 | /13 | P | |
| **53** | | | | | | | | | | 17099.0 | 6 | RM | /31 | | /5 | | new | |
| **54** | | 5 849.80 | | 0.09 | 0.048 | <0.95 | 1 | 17089.9 | 2.9 | 17091.0 | -1 | | 15/12 | | /5 | /12 | P | |
| **55** | | | | | | | | | | 17100.0 | -10 | 1 | 3/1 | | 2 $^3P_2$ | **He** | M | |
| **56** | | 5 854.50 | | | | | | 17076.1 | | 17076.0 | 0 | | 15/9 | | /5 | /9 | new | |
| **57** | | 5 855.63 | | | | | | 17072.8 | | 17073.0 | -0 | | 15/8 | | /5 | /8 | new | |
| **58** | | | | | | | | | | 17072.0 | 1 | | 14/15 | | /5 | /15 | new | |
| **59** | HeI | 5 876.00 | | | | | | 17013.7 | | 17013.0 | 1 | | 14/10 | | /5 | /10 | new | |
| **60** | | 5 897.73 | | | | | | 16951.0 | | 16959.0 | -8 | | 13/12 | | /5 | /12 | new | |
| **61** | W | 5 900.40 | | 0.2 | 0.013 | | 0.7 | 16943.3 | 2.0 | 16945.0 | -2 | | 13/11 | | /5 | /11 | P | |
| **62** | u | 5 904.60 | | 0.2 | 0.009 | | 0.7 | 16931.3 | 2.0 | 16933.0 | -2 | | 13/10 | | /5 | /10 | P | |
| **63** | | 5 908.40 | | 0.2 | 0.003 | | 1.2 | 16920.4 | 3.4 | 16924.0 | -4 | | 13/9 | | /5 | /9 | P | |
| **64** | | 5 910.54 | | 0.2 | 0.015 | | 0.9 | 16914.2 | 2.6 | 16913.0 | 1 | | 13/7 | | /5 | /7 | new | |
| **65** | | 5 922.25 | | | | | | 16880.8 | | | | | | | | | | |
| **66** | | 5 923.40 | | 0.2 | 0.027 | | 0.6 | 16877.5 | 1.7 | 16876.0 | 2 | | 12/12 | | /5 | /12 | new | |
| **67** | | 5 925.90 | | 0.2 | 0.015 | | 1 | 16870.4 | 2.8 | 16874.0 | -4 | | 9/12 | 1 | 2/3 | | new | |
| **68** | | 5 927.68 | | 0.2 | 0.011 | | 0.7 | 16865.3 | 2.0 | 16861.0 | 4 | 1 | 2/4 | | 2 $^3P_2$ | **He** | M | |
| **69** | | 5 945.47 | | 0.2 | 0.013 | | 0.6 | 16814.9 | 1.7 | 16816.0 | -1 | | 11/13 | | /5 | /13 | P | |
| **70** | | 5 947.29 | | 0.2 | 0.017 | | 1 | 16809.7 | 2.8 | 16812.0 | -2 | | 12/8 | | /5 | /8 | P | |
| **71** | | 5 948.86 | | 0.2 | 0.005 | | 0.6 | 16805.3 | 1.7 | 16804.0 | 1 | | 12/7 | | /5 | /7 | P | |
| **72** | W | 5 958.90 | | | | | | 16777.0 | | 16779.0 | -2 | | 11/12 | | /5 | /12 | P | |
| **73** | | 5 963.72 | | | | | | 16763.4 | | | | | | | | | | |
| **74** | | 5 966.71 | | | | | | 16755.0 | | 16752.0 | 3 | | 8/9 | 3 | 2/3 | | new | |
| **75** | | 5 970.25 | | | | | | 16745.1 | | 16747.0 | -2 | | 11/11 | | /5 | /11 | P | |
| **76** | | 5 973.75 | | | | | | 16735.3 | | 16733.0 | 2 | | 9/10 | 1 | 2/3 | | new | |
| **77** | | 5 975.74 | | 0.2 | 0.006 | | 0.8 | 16729.7 | 2.2 | 16727.0 | 3 | 3 | 2/4 | | 2 $^3P_2$ | **He** | M | |
| **78** | u | 5 982.93 | | 0.32 | 0.015 | | 1 | 16709.6 | 2.8 | 16719.0 | -9 | | 11/10 | | /5 | /10 | P | |
| **79** | | 5 986.66 | | 0.2 | 0.008 | | 0.9 | 16699.2 | 2.5 | 16696.0 | 3 | | 9/14 | | /5 | /14 | new | |
| **80** | | 5 988.08 | | 0.2 | 0.012 | | 1.1 | 16695.2 | 3.1 | 16696.0 | -1 | | 11/9 | | /5 | /9 | P | |
| **81** | | 5 989.44 | | | | | | 16691.4 | | 16690.0 | 1 | 1 | 3/2 | | 2 $^2$P | **Li** | | |
| **82** | | 5 995.75 | | 0.2 | 0.012 | | 1.2 | 16673.9 | 3.3 | 16679.0 | -5 | | 11/8 | | /5 | /8 | P | |
| **83** | | 5 999.63 | | 0.2 | 0.006 | | 0.6 | 16663.1 | 1.7 | 16667.0 | -4 | | 10/12 | | /5 | /12 | P | |

| A | A | B | C | D | E | F | G | H | I | J | K | L | M | N | O | P | Q | R |
|---|---|---|---|---|---|---|---|---|---|---|---|---|---|---|---|---|---|---|
| **84** | | 6 005.03 | | 0.55 | 0.024 | | 1.2 | 16648.1 | 3.3 | 16649.0 | -1 | | 11/1 | | /5 | /1 | P | |
| **85** | | 6 010.65 | | 0.11 | 0.141 | | 3.5 | 16632.5 | 9.7 | 16649.0 | -16 | | 11/1 | | /5 | /1 | P | |
| **86** | | 6 019.36 | | 0.1 | 0.018 | | 0.9 | 16608.5 | 2.5 | 16615.0 | -7 | | 8/14 | | /5 | /14 | new | |
| **87** | | 6 027.48 | | 0.2 | 0.029 | | 1.6 | 16586.1 | 4.4 | 16584.0 | 2 | | 9/7 | 1 | 2/3 | | new | |
| **88** | | 6 030.40 | | | | | | 16578.1 | | 16577.0 | 1 | | 10/10 | | /5 | /10 | P | |
| **89** | | 6 032.84 | | 0.6 | 0.005 | | 0.6 | 16571.4 | 1.6 | | | | | | | | | |
| **90** | | 6 037.61 | | 0.23 | 0.039 | | 1.8 | 16558.3 | 4.9 | 16555.0 | 3 | 3,1 | 3/3 | | /4 | /3 | P | |
| **91** | | 6 059.67 | | 0.67 | 0.019 | | 1.2 | 16498.0 | 3.3 | 16511.0 | -13 | | 10/8 | | /5 | /8 | new | |
| **92** | | 6 065.20 | | 0.04 | 0.014 | | 0.94 | 16482.9 | 2.6 | 16490.0 | -7 | | 10/7 | | /5 | /7 | P | |
| **93** | | 6 068.33 | | | | | | 16474.4 | | 16476.0 | -2 | | 10/6 | | /5 | /6 | new | |
| **94** | | 6 084.75 | | | | | | 16430.0 | | 16438.0 | -8 | | 8/9 | 1 | 2/3 | | new | |
| **95** | | 6 089.78 | | 0.08 | 0.017 | | 0.82 | 16416.4 | 2.2 | 16422.0 | -6 | | 7/13 | | /5 | /13 | P | |
| **96** | | | | | | | | | | 16412.0 | 4 | 1 | 2/5 | 3 | 2/2 | | new | |
| **97** | | 6 096.27 | | | | | | 16398.9 | | 16408.0 | -9 | | 9/10 | | /5 | /10 | | |
| **98** | | 6 102.38 | | | | | | 16382.5 | | 16376.0 | 7 | 3 | 2/5 | 3 | 2/2 | | new | |
| **99** | | 6 108.05 | | 0.19 | 0.007 | | 0.58 | 16367.3 | 1.6 | | | | | | | | | |
| **100** | | 6 113.20 | | 0.17 | 0.027 | | 1 | 16353.5 | 2.7 | 16354.0 | -0 | 1 | 3/3 | 1 | 2/1 | | M | |
| **101** | | 6 116.80 | | 0.08 | 0.007 | | 0.52 | 16343.9 | 1.4 | 16347.0 | -3 | | 9/9 | | /5 | /9 | P | |
| **102** | | | | | | | | | | 16344.0 | 0 | 3 | 3/3 | 1 | 2/1 | | M | |
| **103** | | 6 118.68 | | | | | | 16338.9 | | 16334.0 | 5 | 3 | 3/1 | | 2 $^2$P | **Li** | new | |
| **104** | | 6 139.94 | | 0.03 | 0.014 | | 0.88 | 16282.3 | 2.3 | 16285.0 | 3 | 1 | 3/1 | | 2 $^1$P | **He** | M | |
| **105** | | 6 141.91 | | | | | | 16277.1 | | 16281.0 | -4 | | 8/7 | 1 | +2/3 | | new | |
| **106** | | 6 145.69 | | | | | | 16267.1 | | 16260.0 | 7 | | 9/7 | | /5 | /7 | new | |
| **107** | | 6 158.54 | | | | | | 16233.1 | | 16234.0 | -1 | | 9/6 | | /5 | /6 | new | |
| **108** | | 6 161.93 | | | | | | 16224.2 | | 16217.0 | 7 | | 9/5 | | /5 | /5 | new | |
| **109** | | 6 165.97 | | | | | | 16213.6 | | 16213.0 | 1 | | 8/10 | | /5 | /10 | new | |
| **110** | | 6 167.84 | | | | | | 16208.6 | | 16208.0 | 1 | | 9/4 | | /5 | /4 | new | |
| **111** | | 6 185.81 | | | | | | 16161.6 | | 16165.0 | -3 | | 7/9 | 1 | 2/3 | | new | |
| **112** | | 6 194.73 | | 0.34 | 0.006 | | 0.59 | 16138.3 | 1.5 | 16149.0 | -11 | 1 | 3/4 | 3 | 2/2 | | M | |
| **113** | | 6 195.96 | | 0.33 | 0.061 | <0.95 | 0.65 | 16135.1 | 1.7 | 16139.0 | -4 | 3 | 3/4 | 3 | 2/2 | | M | |
| **114** | | 6 198.87 | | | | | | 16127.5 | | | | | | | | | | |
| **115** | br | 6 203.08 | | 0.38 | 0.107 | <0.95 | 1.2 | 16116.6 | 3.1 | 16113.5 | -1 | | 8/9 | | /5 | /9 | P | |
| **116** | | | | | | | | | | 16117.0 | 0 | 0 | 3/2 | 3 | 2/1 | | new | |
| **117** | | 6 204.66 | | 0.3 | 0.189 | | 3.9 | 16112.5 | 10.1 | 16117 | -5 | 0 | 3/2 | 3 | 2/1 | | new | |
| **118** | W? | 6 211.67 | | 0.4 | 0.014 | | 1.2 | 16094.3 | 3.1 | 16097.0 | -3 | 3 | 3/2 | 3 | 2/1 | | M | |
| **119** | W? | 6 212.90 | | | | | | 16091.1 | | | | | | | | | | |
| **120** | | 6 215.79 | | 0.52 | 0.007 | | 1.1 | 16083.6 | 2.8 | 16086.0 | -2 | 3 | 2/4 | | 2 $^2$P | **Li** | new | |
| **121** | | 6 220.81 | | 0.2 | 0.006 | | 0.8 | 16070.6 | 2.1 | | | | | | | | | |
| **122** | | 6 223.56 | | 0.43 | 0.005 | | 0.54 | 16063.5 | 1.4 | 16059.0 | 5 | | 4 $^2P_{3/2}$ | 1 | 2/1 | **Na** | new | |
| **123** | | 6 226.30 | | | | | | 16056.5 | | 16053.0 | 3 | | 4 $^2P_{1/2}$ | 1 | 2/1 | **Na** | new | |
| **124** | | 6 234.30 | | 0.36 | 0.022 | | 0.83 | 16035.9 | 2.1 | | | | | | | | | |
| **125** | | 6 236.67 | | 0.81 | 0.007 | | 0.57 | 16029.8 | 1.5 | 16027.0 | 3 | | 8/8 | | /5 | /8 | P | |
| **126** | | 6 250.84 | | | | | | 15993.4 | | 15998.0 | -5 | | 7/10 | | /5 | /10 | new | |
| **127** | | 6 269.75 | | 0.33 | 0.076 | <0.95 | 1 | 15945.2 | 2.5 | 15951.0 | -6 | 1 | 3/1 | 3 | 2/1 | | M | |
| **128** | | | | | | | | | | 15944.0 | 1 | 1 | 2/5 | 1 | 2/2 | | new | |
| **129** | STR | 6 283.85 | | 0.33 | 0.618 | <0.95 | 2.6 | 15909.4 | 6.6 | 15908.0 | 1 | 3 | 2/5 | 1 | 2/2 | | new | |
| **130** | | 6 287.47 | | 0.41 | 0.014 | | 0.62 | 15900.3 | 1.6 | 15906.0 | -6 | | 8/6 | | /5 | /6 | new | |
| **131** | | 6 289.74 | | 0.81 | 0.012 | | 1.1 | 15894.5 | 2.8 | 15884.0 | 11 | 3 | 3/1 | | /6 | /1 | new | |
| **132** | | 6 309.10 | | 0.25 | 0.058 | | 2.4 | 15845.7 | 6.0 | 15846.0 | -0 | | 8/3 | | /5 | /3 | new | |
| **133** | | 6 315.96 | | 2.1 | | | 2 | 15828.5 | 5.0 | 15840.0 | -11 | | 8/1 | | /5 | /1 | P | |
| **134** | | 6 317.06 | | 0.2 | 0.058 | | 2.1 | 15825.8 | 5.3 | 15825.0 | 1 | 3 | 3/1 | 3 | 2/1 | **He** | M | |
| **135** | | 6 318.30 | | | | | | 15822.7 | | | | | | | | | | |
| **136** | | 6 320.54 | | 0.44 | 0.016 | | 1.5 | 15817.1 | 3.8 | | | | | | | | | |
| **137** | | 6 324.80 | | 0.25 | 0.018 | | 0.83 | 15806.4 | 2.1 | | | | | | | | | |
| **138** | | 6 329.97 | | 0.19 | 0.018 | | 1 | 15793.5 | 2.5 | 15799.0 | -6 | | 3 $^2$D | 3 | 2/1 | **Na** | new | |
| **139** | | 6 353.34 | | 0.24 | 0.036 | | 1.8 | 15735.4 | 4.5 | | | | | | | | | |
| **140** | | 6 362.30 | | 1.27 | 0.025 | | 1.3 | 15713.2 | 3.2 | 15 712.0 | 1 | 1 | 2/4 | 3 | 2/1 | | M | |
| **141** | | 6 367.25 | | 0.07 | 0.017 | | 0.88 | 15701.0 | 2.2 | 15700.0 | 1 | 1 | 2/5 | | 3 $^2P_{3/2}$ | **Na** | new | |
| **142** | | 6 368.58 | | 0.2 | 0.013 | | 0.8 | 15697.7 | 2.0 | 15690.0 | 8 | | 7/8 | | /5 | /8 | new | |
| **143** | | 6 375.95 | | 0.03 | 0.026 | <0.95 | 0.74 | 15679.6 | 1.8 | 15681.0 | -1 | 1 | 3/4 | 1 | 2/2 | | M | |
| **144** | | 6 377.14 | | 0.12 | 0.032 | | 3.5 | 15676.7 | 8.6 | 15676 | 1 | 0 | 3/4 | 1 | 2/2 | | new | |
| **145** | | 6 379.29 | | 0.08 | 0.078 | <0.95 | 0.79 | 15671.4 | 1.9 | 15671.0 | 0 | 3 | 3/4 | 1 | 2/2 | | M | |
| **146** | | 6 397.39 | | 0.27 | 0.025 | | 1.4 | 15627.1 | 3.4 | | | | | | | | | |
| **147** | | 6 400.30 | | 0.2 | 0.003 | | 0.8 | 15619.9 | 2.0 | 15618.0 | 2 | | 6/8 | 1 | 2/3 | | new | |
| **148** | | 6 410.18 | | | | | | 15595.9 | | | | | | | | | | |
| **149** | | 6 413.93 | | 0.29 | 0.016 | | 0.88 | 15586.8 | 2.1 | 15578.0 | 9 | 3 | 2/4 | 3 | 2/1 | | M | |
| **150** | | 6 425.70 | | 0.2 | 0.019 | | 0.71 | 15558.2 | 1.7 | 15559.0 | -1 | | 7/7 | | /5 | /7 | new | |
| **151** | | 6 439.50 | | 0.24 | 0.02 | | 1 | 15524.9 | 2.4 | | | | | | | | | |
| **152** | | 6 445.20 | | 0.36 | 0.048 | | 1.2 | 15511.1 | 2.9 | 15513.0 | -2 | | 3 $^2$P | 3 | 2/1 | **Li** | new | |
| **153** | | 6 449.14 | | 0.27 | 0.02 | | 1.2 | 15501.7 | 2.9 | 15501.0 | 1 | 0 | 3/1 | | /4 | /1 | new | |
| **154** | | 6 460.00 | | 0.2 | 0.009 | | 0.79 | 15475.6 | 1.9 | 15475.0 | 1 | | 6/6 | 3 | 2/3 | | new | |
| **155** | | 6 463.61 | | | | | | 15467.0 | | | | | | | | | | |
| **156** | | 6 465.48 | | | | | | 15462.5 | | | | | | | | | | |
| **157** | | 6 466.74 | | | | | | 15459.5 | | | | | | | | | | |
| **158** | | 6 468.70 | | | | | | 15454.8 | | 15456.0 | -1 | | 7/6 | | /5 | /6 | new | |
| **159** | | 6 474.27 | | | | | | 15441.5 | | 15444.0 | -3 | 3 | 3/4 | | 3 $^2P_{1/2}$ | **Na** | new | |
| **160** | | 6 476.94 | | | | | | 15435.1 | | 15437.0 | -2 | 1 | 3/4 | | 3 $^2P_{3/2}$ | **Na** | new | |
| **161** | | 6 489.38 | | | | | | 15405.5 | | | | | | | | | | |
| **162** | | 6 492.02 | | 0.15 | 0.018 | | 0.76 | 15399.3 | 1.8 | 15387.0 | 12 | | 7/5 | | /5 | /5 | P | |
| **163** | | 6 520.56 | | 0.24 | 0.024 | | 0.97 | 15331.9 | 2.3 | 15342.0 | -10 | 1 | 3/2 | 1 | 2/1 | | M | |
| **164** | | | | | | | | | | 15329.0 | 3 | | 3 $^3$P | 3 | 2/1 | **He** | new | |
| **165** | | 6 543.20 | | | | | | 15278.8 | | 15283.0 | -4 | 1 | 3/2 | | 2 $^1$P | **He** | M | |

| A | A | B | C | D | E | F | G | H | I | J | K | L | M | N | O | P | Q | R |
|---|---|---|---|---|---|---|---|---|---|---|---|---|---|---|---|---|---|---|
| **166** | | 6 546.57 | | | | | | 15271.0 | | | | | | | | | | |
| **167** | | 6 549.07 | | | | | | 15265.1 | | | | | | | | | | |
| **168** | | 6 553.82 | | | | | | 15254.1 | | 15257.0 | -3 | 3 | 3/2 | 1 | 2/1 | | M | |
| **169** | | 6 567.63 | | | | | | 15222.0 | | 15218.0 | 4 | 0 | 3/2 | | 2 $^1$P | **He** | new | |
| **170** | | 6 570.52 | | | | | | 15215.3 | | 15214.0 | 1 | | 6/4 | 3 | 2/3 | | new | |
| **171** | | 6 572.84 | | | | | | 15209.9 | | 15198.0 | 12 | 3 | 3/2 | | 2 $^1$P | **He** | M | |
| **172** | | 6 594.13 | | 0.53 | 0.087 | | 5.6 | 15160.8 | 12.9 | 15161.0 | -0 | | 6/6 | 1 | 2/3 | | new | |
| **173** | | 6 597.31 | | 0.05 | 0.019 | | 0.72 | 15153.5 | 1.7 | 15150.0 | 4 | | 6/2 | 3 | 2/3 | | new | |
| **174** | STR | 6 613.56 | | 0.12 | 0.231 | <0.95 | 1.1 | 15116.3 | 2.5 | 15111.0 | 5 | 1 | 3/1 | 1 | 2/1 | | M | |
| **175** | | 6 630.80 | | | | | | 15077.0 | | 15080.0 | -3 | | 5/7 | 3 | 2/3 | | new | |
| **176** | | 6 631.66 | | | | | | 15075.0 | | 15072.0 | 3 | | 48 | 3 | 2/3 | | new | |
| **177** | | 6 632.85 | | 0.1 | 0.017 | | 1.1 | 15072.3 | 2.5 | 15074.0 | -2 | Ry | | | 3 $^3$S | **He** | new | |
| **178** | | 6 646.03 | | | | | | 15042.4 | | 15052.0 | -10 | 1 | 3/1 | | 2 $^1$P | **He** | M | |
| **179** | | 6 654.58 | | | | | | 15023.1 | | 15029.3 | -6 | | 3 $^3$D$_3$ | 1 | 2/1 | **He** | new | |
| **180** | | 6 660.64 | | 0.13 | 0.051 | <0.95 | 0.84 | 15009.4 | 1.9 | 15026.0 | -17 | | 3 $^3$D$_3$ | 1 | 2/1 | **He** | M | |
| **181** | | 6 665.15 | | | | | | 14999.3 | | 15004.0 | -5 | | 6/5 | 1 | 2/3 | | new | |
| **182** | | 6 672.15 | | | | | | 14983.5 | | 14985.0 | -1 | 3 | 3/1 | 1 | 2/1 | | M | |
| **183** | | 6 682.65 | | | | | | 14960.0 | | 14959.0 | 1 | | 3 $^2$D | 1 | 2/1 | **Na** | new | |
| **184** | | 6 689.30 | | 0.2 | 0.014 | | 0.9 | 14945.1 | 2.0 | | | | | | | | | |
| **185** | | 6 691.72 | | | | | | 14939.7 | | | | | | | | | | |
| **186** | | 6 693.35 | | 0.2 | 0.005 | | 0.8 | 14936.1 | 1.8 | | | | | | | | | |
| **187** | | 6 694.48 | | 0.06 | 0.005 | | 0.64 | 14933.6 | 1.4 | 14934.0 | -0 | 3 | 2/4 | | /4 | /4 | new | |
| **188** | | 6 699.24 | | 0.11 | 0.041 | | 1.2 | 14922.9 | 2.7 | 14926.0 | -3 | 3 | 3/1 | | 2 $^1$P | **He** | M | |
| **189** | | 6 701.98 | | 0.09 | 0.015 | | 0.99 | 14916.8 | 2.2 | | | | | | | | | |
| **190** | | 6 709.39 | | 0.05 | 0.009 | | 1.3 | 14900.4 | 2.9 | 14905.0 | -5 | | 6/4 | 1 | 2/3 | | new | |
| **191** | | 6 729.28 | | | | | | 14856.3 | | 14860.0 | -4 | | 5/8 | | /5 | /8 | new | |
| **192** | | 6 737.13 | | | | | | 14839.0 | | 14837.0 | 2 | | 6/6 | | /5 | /6 | new | |
| **193** | | 6 740.99 | | 0.15 | 0.013 | | 0.97 | 14830.5 | 2.1 | 14832.0 | -1 | | 6/1 | 1 | 2/3 | | new | |
| **194** | | 6 767.74 | | 0.1 | 0.005 | | 0.7 | 14771.9 | 1.5 | | | | | | | | | |
| **195** | | 6 770.05 | | 0.13 | 0.013 | | 0.74 | 14766.9 | 1.6 | 14766.0 | 1 | | 5/7 | 1 | 2/3 | | new | |
| **196** | | 6 778.99 | | 0.1 | 0.003 | | 0.54 | 14747.4 | 1.2 | 14738.0 | 9 | 3 | 2/4 | 1 | 2/1 | | M | |
| **197** | | 6 788.66 | | 0.08 | 0.007 | | 0.87 | 14726.4 | 1.9 | | | | | | | | | |
| **198** | | 6 792.52 | | 0.17 | 0.015 | | 1.1 | 14718.0 | 2.4 | | | | | | | | | |
| **199** | | 6 795.24 | | 0.12 | 0.009 | | 0.9 | 14712.1 | 1.9 | | | | | | | | | |
| **200** | | 6 801.37 | | 0.11 | 0.007 | | 0.66 | 14698.9 | 1.4 | | | | | | | | | |
| **201** | | 6 803.29 | | 0.1 | 0.005 | | 0.8 | 14694.7 | 1.7 | | | | | | | | | |
| **202** | | 6 807.31 | | | | | | 14686.0 | | | | | | | | | | |
| **203** | | 6 810.49 | | 0.2 | 0.018 | <0.95 | 1.5 | 14679.2 | 3.2 | 14679.0 | 0 | | 6/5 | | /5 | /5 | new | |
| **204** | | | | | | | | | | 14679.0 | 0 | 3 | 2/4 | | 2 $^1$P | **He** | M | |
| **205** | | 6 821.56 | | 0.1 | 0.011 | | 1.3 | 14655.4 | 2.8 | | | | | | | | | |
| **206** | | 6 823.30 | | 0.1 | 0.006 | | 0.8 | 14651.6 | 1.7 | 14652.0 | -0 | | 5/6 | 3 | 2/3 | | new | |
| **207** | | 6 827.30 | | 0.06 | 0.015 | | 0.96 | 14643.0 | 2.1 | | | | | | | | | |
| **208** | | 6 834.50 | | 0.1 | 0.007 | | 0.85 | 14627.6 | 1.8 | 14629.1 | -1 | | 3 $^1$S | 3 | 2/1 | **He** | new | |
| **209** | | 6 837.70 | | 0.1 | 0.008 | | 0.67 | 14620.8 | 1.4 | | | | | | | | | |
| **210** | | 6 841.49 | | 0.13 | 0.007 | | 0.62 | 14612.7 | 1.3 | | | | | | | | | |
| **211** | | 6 843.60 | | 0.05 | 0.027 | | 1.2 | 14608.2 | 2.6 | 14610.0 | -2 | | 15/12 | | 3 $^3$S | **He** | new | |
| **212** | | 6 845.30 | | 0.1 | 0.006 | | 0.73 | 14604.5 | 1.6 | | | | | | | | | |
| **213** | | 6 846.60 | | 0.1 | 0.005 | | 0.56 | 14601.8 | 1.2 | 14603.0 | -1 | | 15/11 | | 3 $^3$S | **He** | new | |
| **214** | | 6 847.76 | | 0.1 | 0.005 | | 0.7 | 14599.3 | 1.5 | 14598.0 | 1 | | 15/10 | | 3 $^3$S | **He** | new | |
| **215** | | 6 849.56 | | | | | | 14595.5 | | 14595.0 | 0 | | 15/9 | | 3 $^3$S | **He** | new | |
| **216** | | 6 852.67 | | 0.19 | 0.018 | | 1.4 | 14588.8 | 3.0 | | | | | | | | | |
| **217** | | 6 855.38 | | | | | | 14583.1 | | 14581.0 | 2 | | 6/4 | | /5 | /4 | new | |
| **218** | | 6 860.02 | | 0.14 | 0.024 | | 0.93 | 14573.2 | 2.0 | 14575.0 | -2 | | 14/13 | | 3 $^3$S | **He** | new | |
| **219** | | 6 862.53 | | 0.1 | 0.02 | | 0.9 | 14567.9 | 1.9 | | | | | | | | | |
| **220** | | 6 864.65 | | | | | | 14563.4 | | 14561.0 | 2 | | 14/13 | | 3 $^3$S | **He** | new | |
| **221** | | 6 886.56 | | 0.1 | 0.038 | | 0.73 | 14517.0 | 1.5 | 14513.0 | 4 | | 6/2 | | /5 | /2 | new | |
| **222** | | 6 919.44 | | 0.04 | 0.053 | | 0.96 | 14448.1 | 2.0 | 14442.0 | 6 | | 5/7 | | /5 | /7 | new | |
| **223** | | 6 944.56 | | 0.08 | 0.028 | | 0.84 | 14395.8 | 1.7 | 14394.0 | 2 | | 12/12 | | 3 $^3$S | **He** | new | |
| **224** | | 6 950.55 | | | | | | 14383.4 | | | | | | | | | | |
| **225** | | 6 953.60 | | | | | | 14377.1 | | 14374.0 | 3 | | 11/14 | | 3 $^3$S | **He** | new | |
| **226** | | 6 963.54 | | | | | | 14356.6 | | 14355.0 | 2 | | 12/10 | | 3 $^3$S | **He** | new | |
| **227** | | 6 971.51 | | | | | | 14340.1 | | 14341.0 | -1 | | 12/9 | | 3 $^3$S | **He** | new | |
| **228** | | 6 973.55 | | | | | | 14335.9 | | 14338.0 | -2 | | 5/6 | 1 | 2/3 | | new | |
| **229** | | 6 978.28 | | 0.06 | 0.012 | | 0.81 | 14326.2 | 1.7 | 14323.0 | 3 | | 12/7 | | 3 $^3$S | **He** | new | |
| **230** | | 6 982.46 | | | | | | 14317.7 | | 14318.0 | -0 | | 12/6 | | 3 $^3$S | **He** | new | |
| **231** | | 6 993.18 | | 0.09 | 0.116 | <0.95 | 0.96 | 14295.7 | 2.0 | 14297.0 | -1 | 3 | 3/3 | 3 | 2/2 | | M | |
| **232** | | 6 998.76 | | 0.11 | 0.016 | | 0.56 | 14284.3 | 1.1 | | | | | | | | | |
| **233** | | 7 002.19 | | | | | | 14277.3 | | 14279.0 | -2 | | 5/5 | 3 | 2/3 | | new | |
| **234** | | 7 030.35 | | | | | | 14220.1 | | | | | | | | | | |
| **235** | | 7 031.56 | | | | | | 14217.7 | | 14215.0 | 3 | | 11/9 | | 3 $^3$S | **He** | new | |
| **236** | | 7 032.88 | | | | | | 14215.0 | | 14214.0 | 1 | | 9/14 | | 3 $^3$S | **He** | new | |
| **237** | | 7 045.87 | | 0.25 | 0.013 | | 0.84 | 14188.8 | 1.7 | 14189.0 | -0 | | 10/12 | | 3 $^3$S | **He** | new | |
| **238** | | 7 060.05 | | 0.32 | 0.019 | | 0.67 | 14160.3 | 1.3 | | | | | | | | | |
| **239** | | 7 061.00 | | | | | | 14158.4 | | | | | | | | | | |
| **240** | | 7 062.65 | | 0.1 | 0.023 | | 0.6 | 14155.1 | 1.2 | 14153.0 | 2 | 1 | 2/3 | | 2 $^1$S | **He** | M | |
| **241** | | 7 065.57 | | | | | | 14149.2 | | | | | | | | | | |
| **242** | | 7 069.48 | | 0.19 | 0.023 | | 0.92 | 14141.4 | 1.8 | 14142.0 | -1 | | 9/13 | | 3 $^3$S | **He** | new | |
| **243** | | 7 072.66 | | | | | | 14135.1 | | 14139.0 | -4 | | 10/11 | | 3 $^3$S | **He** | new | |
| **244** | | 7 078.11 | | 0.2 | 0.013 | | 0.71 | 14124.2 | 1.4 | | | | | | | | | |
| **245** | | 7 084.94 | | 0.13 | 0.03 | | 2 | 14110.6 | 4.0 | 14116.0 | -5 | | 4/7 | 1 | 2/3 | | new | |
| **246** | br | 7 105.81 | | 0.34 | 0.042 | | 2.5 | 14069.1 | 5.0 | 14068.0 | 1 | | 9/12 | | 3 $^3$S | **He** | new | |
| **247** | | 7 137.77 | | 0.51 | 0.045 | | 3.5 | 14006.1 | 6.9 | 14009.0 | -3 | | 10/7 | | 3 $^3$S | **He** | new | |
| **248** | | 7 153.94 | | 0.17 | 0.007 | | 0.67 | 13974.5 | 1.3 | 13977.0 | -3 | | 10/1 | | 3 $^3$S | **He** | new | |

| A | A | B | C | D | E | F | G | H | I | J | K | L | M | N | O | P | Q | R |
|---|---|---|---|---|---|---|---|---|---|---|---|---|---|---|---|---|---|---|
| **249** | | 7 159.51 | | | | | | 13963.6 | | | | | | | | | | |
| **250** | | 7 161.30 | | 0.23 | 0.059 | | 2.2 | 13960.1 | 4.3 | 13965.0 | -5 | | 5/5 | 1 | 2/3 | | new | |
| **251** | | 7 162.96 | | | | | | 13956.9 | | | | | | | | | | |
| **252** | STR | 7 224.00 | | 0.21 | 0.259 | <0.95 | 1.1 | 13838.9 | 2.1 | 13839.0 | -0 | 3 | 2/3 | | 2 $^{1}$S | **He** | M | |
| **253** | | | | | | | | | | 13839.0 | 0 | 1 | 3/3 | 1 | 2/2 | | M | |
| **254** | | 7 228.49 | | 0.18 | 0.026 | | 0.89 | 13830.3 | 1.7 | 13829.0 | 1 | 3 | 3/3 | 1 | 2/2 | | M | |
| **255** | | 7 249.28 | | 0.3 | 0.055 | | 1.3 | 13790.7 | 2.5 | 13789.1 | 2 | | 3 $^{1}$S | 1 | 2/1 | **He** | new | |
| **256** | | 7 257.49 | | 0.3 | 0.019 | | 0.61 | 13775.1 | 1.2 | 13778.0 | -3 | | 9/7 | | 3 $^{3}$S | **He** | new | |
| **257** | | 7 267.95 | | 0.47 | 0.029 | | 1.7 | 13755.2 | 3.2 | 13752.0 | 3 | | 9/6 | | 3 $^{3}$S | **He** | new | |
| **258** | | 7 349.79 | | 0.12 | 0.013 | | 0.84 | 13602.1 | 1.6 | 13602.0 | 0 | 3 | 3/3 | | 3 $^{2}$P$_{1/2}$ | **Na** | new | |
| **259** | | 7 354.60 | | 0.26 | 0.008 | | 0.64 | 13593.2 | 1.2 | 13595.0 | -2 | 1 | 3/3 | | 3 $^{2}$P$_{3/2}$ | **Na** | new | |
| **260** | | 7 357.60 | | 1.81 | 0.048 | | 1.4 | 13587.6 | 2.6 | | | | | | | | | |
| **261** | | 7 360.49 | | 0.23 | 0.021 | | 1 | 13582.3 | 1.8 | 13585.0 | -3 | 3 | 3/3 | | 3 $^{2}$P$_{3/2}$ | **Na** | new | |
| **262** | | 7 367.12 | | 0.12 | 0.042 | | 1.3 | 13570.1 | 2.4 | | | | | | | | | |
| **263** | | 7 369.94 | | 0.51 | 0.019 | | 1 | 13564.9 | 1.8 | 13560.0 | 5 | | 5/3 | 1 | 2/3 | | new | |
| **264** | | 7 375.90 | | 0.15 | 0.013 | | 0.79 | 13553.9 | 1.5 | | | | | | | | | |
| **265** | | 7 385.83 | | 0.08 | 0.009 | | 0.54 | 13535.7 | 1.0 | 13538.0 | -2 | | 4 $^{2}$P$_{1/2}$ | 1 | 2/2 | **Na** | new | |
| **266** | | 7 406.30 | | 0.16 | 0.014 | | 1.2 | 13498.3 | 2.2 | | | | | | | | | |
| **267** | | 7 419.07 | | | | | | 13475.1 | | 13475.0 | 0 | | 8/7 | | 3 $^{3}$S | **He** | new | |
| **268** | | 7 458.15 | | | | | | 13404.5 | | | | | | | | | | |
| **269** | | 7 468.90 | | | | | | 13385.2 | | | | | | | | | | |
| **270** | | 7 470.35 | | | | | | 13382.6 | | | | | | | | | | |
| **271** | | 7 472.65 | | | | | | 13378.4 | | 13379.0 | -1 | | 5/4 | | /5 | /4 | new | |
| **272** | | 7 483.02 | | | | | | 13359.9 | | 13360.0 | -0 | | 8/2 | | 3 $^{3}$S | **He** | new | |
| **273** | | 7 484.09 | | | | | | 13358.0 | | 13359.0 | -1 | | 7/9 | | 3 $^{3}$S | **He** | new | |
| **274** | | 7 494.89 | | | | | | 13338.8 | | 13347.0 | -8 | | 4/6 | 1 | 2/3 | | new | |
| **275** | | 7 559.35 | | 0.6 | 0.034 | | 1.5 | 13225.0 | 2.6 | | | | | | | | | |
| **276** | | 7 562.00 | | 0.2 | 0.087 | <0.95 | | 13220.4 | | 13210.0 | 10 | 3 | 3/2 | 3 | 2/2 | | M | |
| **277** | | 7 571.66 | | | | | 1.78 | 13203.5 | 3.1 | 13208.0 | -4 | | 7/8 | | 3 $^{3}$S | **He** | new | |
| **278** | | 7 580.05 | | | | | | 13188.9 | | | | | | | | | | |
| **279** | | 7 581.30 | | 0.13 | 0.036 | | 1.5 | 13186.7 | 2.6 | | | | | | | | | |
| **280** | | 7 696.00 | | 0.16 | 0.009 | | 0.67 | 12990.2 | 1.1 | 13001.0 | -11 | | 3 $^{3}$S | 3 | 2/1 | **He** | new | |
| **281** | KI | 7 699.00 | | | | <0.95 | | 12985.1 | | 12979.0 | 6 | | 3 $^{3}$D$_{3}$ | 3 | 2/2 | **He** | M | |
| **282** | | 7 707.96 | | | | | | 12970.0 | | 12974.0 | -4 | | 7/6 | | 3 $^{3}$S | **He** | new | |
| **283** | | 7 720.26 | | | | | | 12949.4 | | | | | | | | | | |
| **284** | | 7 721.85 | | 0.17 | 0.023 | | 0.79 | 12946.7 | 1.3 | 12938.0 | 9 | 3 | 3/1 | 3 | 2/2 | | M | |
| **285** | | 7 832.81 | | 0.06 | 0.023 | | 0.84 | 12763.3 | 1.4 | 12762.0 | 1 | 1 | 2/2 | 0 | 3/2 | | new | |
| **286** | | 7 862.39 | | 0.1 | 0.009 | | 0.67 | 12715.3 | 1.1 | | | | | | | | | |
| **287** | | 7 908.75 | | | | | | 12640.7 | | | | | | | | | | |
| **288** | br | 7 915.36 | | 0.48 | 0.019 | | 1.9 | 12630.2 | 3.0 | 12644.0 | -14 | | 3/6 | 3 | 2/3 | | new | |
| **289** | | 7 920.70 | | | | | | 12621.7 | | 12626.0 | -4 | | 3 $^{2}$P | 3 | 2/2 | **Li** | new | |
| **290** | | 8 026.27 | | 0.15 | 0.042 | <0.95 | 0.79 | 12455.7 | 1.2 | 12470.0 | -14 | 3 | 3/1 | 1 | 2/2 | | M | |
| **291** | | 8 037.90 | | 0.23 | 0.037 | | 1.7 | 12437.6 | 2.6 | 12444.0 | -6 | Ry | | 3 | 2/4 | | M | |
| **292** | | 8 125.75 | | | | | | 12303.2 | | 12310.0 | -7 | Ry | | 1 | 2/4 | | M | |
| **293** | | 8 283.29 | | 0.18 | 0.028 | | 1.2 | 12069.2 | 1.7 | 12071.0 | -2 | Ry | | 1 | 3/1 | | M | |
| **294** | | 8 439.38 | | | | <0.95 | | 11846.0 | | 11840.0 | 6 | Ry | | 1 | 3/2 | | M | |
| **295** | br | 8 620.79 | | 0.34 | 0.125 | <0.95 | 1.9 | 11596.7 | 2.6 | 11595.0 | 2 | 1 | 4/3 | 3 | 2/3 | | new | |
| **296** | | | | | | | | | | | | | | | | | | |
| **297** | | | | | | | | | | | | | | | | | | |
| **298** | Comments | | | Cited error from JD | | | FWHM in Å | | | | | Upper state | | Lower state | | | Refs. | |
| **299** | from GRMW | | | | Equivalent width from JD | | | Corrected wavenumbers | | | | | | | | | M: Holmlid 2008a | |
| **300** | | | | | | Absorption from GRMW | | | | Calculated wavenumbers | | | | | | | P: Holmlid 2004a | |
| **301** | | | | | | | | | | | difference | | | | | | | |
| **302** | | | | | | | | | FWHM in cm-1 | | | | | | | Atom or second | | |
| **303** | | | | GRMW: Galazutdinov et al. (2000) | | | | | | | | | | | | low atom | | |
| **304** | | | | JD: Jenniskens and Désert (1994) | | | | | | | | | | | | | | |